# Electronic Quantum Coherence in Glycine Molecules Probed with Ultrashort X-ray Pulses in Real Time

David Schwickert[1], Marco Ruberti[2], Přemysl Kolorenč[3], Sergey Usenko[1,4,†], Andreas Przystawik[1], Karolin Baev[1,4], Ivan Baev[4], Markus Braune[1], Lars Bocklage[1,5], Marie Kristin Czwalinna[1], Sascha Deinert[1], Stefan Düsterer[1], Andreas Hans[6], Gregor Hartmann[6], Christian Haunhorst[7], Marion Kuhlmann[1], Steffen Palutke[1], Ralf Röhlsberger[1,5], Juliane Rönsch-Schulenburg[1], Philipp Schmidt[6], Sven Toleikis[1], Jens Viefhaus[8], Michael Martins[4], André Knie[6], Detlef Kip[7], Vitali Averbukh[2], Jon P. Marangos[2], Tim Laarmann[1,5*]

*[1]Deutsches Elektronen-Synchrotron DESY, Notkestr. 85, 22607 Hamburg, Germany*

*[2]Department of Physics, Imperial College London, Prince Consort Road, London SW7 2AZ, United Kingdom*

*[3]Charles University, Faculty of Mathematics and Physics, V Holesovickach 2, 180 00 Praha 8, Czech Republic*

*[4]Department of Physics, University of Hamburg, Luruper Chaussee 149, 22761 Hamburg, Germany*

*[5]The Hamburg Centre for Ultrafast Imaging CUI, Luruper Chaussee 149, 22761 Hamburg, Germany*

*[6]Institute of Physics, University of Kassel, Heinrich-Plett-Str. 40, 34132 Kassel, Germany*

*[7]Faculty of Electrical Engineering, Helmut Schmidt University, Holstenhofweg 85, 22043 Hamburg, Germany*

*[8]Helmholtz-Zentrum Berlin für Materialien und Energie, Albert-Einstein-Straße 15, 12489 Berlin, Germany*

*[†]Present address: European XFEL GmbH, Holzkoppel 4, Schenefeld 22869, Germany*

**[*]Corresponding author. Email: tim.laarmann@desy.de**

Quantum coherence between electronic states of a photoionized molecule and the resulting process of ultrafast electron-hole migration have been put forward as a possible quantum mechanism of charge-directed reactivity governing the photoionization-induced molecular decomposition. Attosecond experiments based on the indirect (fragment ion-based) characterization of the proposed electronic phenomena suggest that the photoionization-induced electronic coherence can survive for tens of femtoseconds, while some theoretical studies predict much faster decay of the coherence due to the quantum uncertainty in the nuclear positions and the nuclear-motion effects. The open questions are: do long-lived electronic quantum coherences exist in complex molecules and can they be probed directly, i.e. via electronic observables? Here, we use x-rays both to create and to directly probe quantum coherence in the photoionized amino acid glycine. The outgoing photoelectron wave leaves behind a positively charged ion that is in a coherent superposition of quantum mechanical eigenstates lying within the ionizing pulse spectral bandwidth. Delayed x-ray pulses track the induced coherence through resonant x-ray absorption that induces Auger decay and by the photoelectron emission from sequential double photoionization. Sinusoidal temporal modulation of the detected signal at early times (0 - 25 fs) is observed in both measurements. Advanced *ab initio* many-electron simulations, taking into account the quantum uncertainty in the nuclear positions, allow us to explain the first 25 fs of the detected coherent quantum evolution in terms of the electronic coherence. Further, in the x-ray absorption measurement we monitor the electronic dynamics for a period of 175 fs and observe an evolving modulation that may implicate the coupling of electronic to vibronic coherence at longer time scales. Our experiment provides a direct support for the existence of long-lived electronic coherence in photoionized biomolecules.

**Introduction**

Understanding the formation and temporal evolution of electronic coherences and the dynamics of coherent superpositions of many-electron configurations forming electronic wave packets that propagate in space and time, is one of the grand challenges in ultrafast science. Electronic coherence may well play a role in photochemical change in matter, in particular in the onset of radiation damage under a wide range of scenarios, from single-molecule x-ray imaging to radiobiology. Its effect in these diverse branches of science, while likely to be present, is yet to be fully understood. Identifying the coherent quantum effects in the radiation damage processes at the molecular level can pave the way to controlling, optimizing and engineering the ionizing radiation damage similarly to how Auger electron-emitting agents are already used in radiotherapy (1), while the much more recently discovered inter-atomic decay phenomena are projected to play a role in fundamental and applied radiobiology (2).

In recent years, intensive debate was stimulated by a seminal study reporting long-lived electronic coherences in polyatomic molecules (3). There, extreme ultraviolet (XUV) pulses ionize the amino acid phenylalanine. A coherent superposition involving several electronic states in the cation is formed due to the broad spectral bandwidth of the sub-femtosecond (fs) XUV pulse. Near-infrared pulses probe the initiated coherent dynamics by photofragmentation. The recorded ion fragmentation pattern shows an oscillatory behavior as a function of pump-probe delay with a time period of 4.3 fs that persists for tens of fs. Matching the corresponding energy difference of relevant energy levels, the results have been interpreted as long-lived electronic coherences. Still, the key interrelated questions are whether the electronic coherence can really be that robust and whether it can be detected directly, i.e. through electronic observables?

The picture of charge migration as purely electronic wave packet dynamics of a coherent superposition of cationic many-electron states (4) only holds within the strict Born-Oppenheimer approximation, where one assumes that the adiabatic separation between the electronic and the nuclear degrees of freedom to be exact. It is widely appreciated, however, that especially in polyatomic molecules of relevance to charge migration, the Born-Oppenheimer approximation breaks down spectacularly, for example due to conical intersections between the adiabatic potential energy surfaces (5). This is clearly relevant, for example, for the regions of partial and full breakdown of the molecular orbital picture of ionization (6), with narrow energy gaps between multiple electronic states of mixed one-hole (1h) and two-hole-one-particle (2h1p) character being on the order of the vibrational quanta. It is, therefore, a relevant open question, to what extent and on what time scale can the original purely electronic picture of charge migration serve as an adequate approximation for the combined electron-nuclear dynamics triggered by molecular ionization? The answer to this question may well be strongly system-dependent.

Recently, the quantum dynamics of nuclear wave packets in phenylalanine has been shown to be prone to destroy electronic coherences within just a few fs (7). Theoretical works treating nuclear dynamics classically found that electron-hole migration in other prototypical model systems, such as tryptophan (8), benzene (9) and glycine (10) is not significantly perturbed by nuclear motion within the first 20 fs after the excitation. In these calculations electronic wave packet decoherence partly stems from the nuclear geometry uncertainty related to the zero-point energy (11), that is also at the core of our present theoretical treatment and partly from the nuclear wave packet dynamics on the potential energy surfaces of the coherently populated electronic states (7). Adding to the controversy, all the experimental evidence on coherence-driven charge migration on the scale of 10s of fs available so far, relied on the indirect detection channels, e.g. ionic fragment detection (3), and no direct measurements (i.e. such based on the electronic observables) of coherent multielectron dynamics have been available.

In the present contribution, we monitor the birth, propagation and fate of an electronic wave packet generated in the interaction of fs x-ray pulses with glycine molecules by means of time-resolved electron spectroscopy at the carbon K-edge from a second fs x-ray probe pulse. The second electron emission in the probe process results from two alternative mechanisms that we have isolated to study them independently of each other. First, a resonant transition can be induced from the carbon core shell into the ionized valence orbital of interest (x-ray absorption), resulting in the Auger decay of the formed core-hole state. Second, a sequential double ionization of the valence-electron shell takes place, a process with weak orbital and spatial selectivity due to the molecular character of the valence orbitals with only moderately orbital-dependent photoionization cross-sections, that nevertheless carries signatures of the temporal evolution of the electronic states of the cation.

The x-ray absorption process, first proposed as a valence-hole coherence probe in (12), involving a localized core orbital and a resonant transition is characterized by high element and orbital selectivity. Thereby, we probe the transient local electron-hole density in inner-valence molecular orbitals directly in the time domain tracing long-lived quantum coherences that result from the initial formation of a coherent superposition of cationic eigenstates in the course of photoionization. Comparison of our experimental results from both channels with high-level *ab initio* many-electron theory taking into account the quantum uncertainty in the nuclear geometry shows that our observations are consistent with the idea of a long-lived electronic coherence.

## Results

Glycine is the simplest of the twenty natural amino acids acting as molecular building blocks of peptides and proteins. It contains the characteristic amine (–NH$_2$) and carboxyl (–COOH) functional groups linked via a methylene bridge and is sketched in the center of Fig. 1B. The neutral molecule has $N = 40$ electrons forming a closed-shell system in its ground state, which can be described by 20 molecular orbitals. A recent theoretical paper reported photoionization cross-sections for the different electron orbitals using density functional theory (13). The theoretical results describing the static electronic structure are in good agreement with experimental photoelectron spectra recorded at a synchrotron light source (14). The valence and inner-valence electron orbitals (6a' - 16a', 1a" - 4a") exhibit binding energies $E_b$ in the range of approximately (10 - 35) eV (10, 13, 14). The Auger spectrum of glycine resulting from oxygen K-shell ionization has been reported in (15).

In our time-resolved experiments we used a single-color, femtosecond x-ray pump-probe technique (16, 17). The electron dynamics were initiated by free-electron laser (FEL) pulses of a few-fs duration at a central photon energy tuned in the range between 269 eV and 281 eV. Whereas the photon energy was sufficient to ionize electrons from all valence- and inner-valence orbitals, the localized 1s core electrons of carbon (~284 eV), nitrogen (~400 eV) and oxygen (~532 eV) were not accessible. The FEL pulses were focused onto the molecular beam produced from glycine powder heated to 160 °C. Electrons generated at the intersection volume were detected with a magnetic bottle electron spectrometer. The instrument also allowed for simultaneous measurement of the resulting ion time-of-flight spectra. More details on the experimental setup can be found in the Materials and Methods section.

We measured the kinetic energy $E_{kin}$ of emitted electrons as a function of FEL photon energy $E_{FEL} = h \cdot \nu$ tuned below the carbon K-edge (18). Fig. 1A shows the recorded kinetic energy distributions, comprising contributions from the initial photoelectron emission, the subsequent Auger decay channel opened by the x-ray probe-pulse excitation of localized carbon 1s electrons and probe-induced photoionization of singly charged glycine cations. Emission bands originating from prompt photoionization of valence- and inner-valence orbitals of the neutral molecule with binding energies $E_b$ shift in their kinetic energy according to $E_{kin} = h \cdot \nu - E_b$. The overall width of the distribution is ~25 eV in agreement with the published data (13, 14). Additional electron emission bands are observed that do not shift as a function of photon energy. These electrons originate from the probe-induced Auger decay, as their kinetic energy is solely given by the electronic structure of glycine molecules. A competing ionization channel resulting in doubly charged glycine ions is the probe-induced photoemission of valence electrons from the initially formed cation. The low-energy onset of their kinetic energy distribution is shifted towards lower energy by the difference between the 1st and the 2nd ionization potential (12 eV) as indicated in Fig. 1A. Note the high-energy tail of the kinetic energy distribution may still reach the upper limit taking into account photoelectron emission from electronically excited cations.

In the following we discuss the overall coherent dynamics tracked in the present time-resolved experiments. The total cascade involves several different processes: (1) photoionization, (2) charge migration and (3a) resonant carbon 1s core-hole excitation followed by Auger decay (x-ray absorption channel) or (3b) sequential double photoionization of glycine depicted in Fig. 1B. The resonance-enhanced, probe-induced Auger decay channel (x-ray absorption) becomes clearly visible by plotting the two-electron coincidences as a function of FEL photon energy with one electron being detected at the kinetic energy corresponding to valence ionization of 10a'. These data are recorded in a second set of measurements with significantly increased data acquisition time by a factor ~100 and are shown in Fig. 1C.

Upon photoionization with fs x-ray pulses we generate electronically excited many-body states in the glycine cation (19). The FEL spectral bandwidth below 0.45% full width at half maximum was sufficient to form a coherent superposition of several cationic eigenstates. Details on the longitudinal coherence and spectral properties of the applied single-mode FEL pulses can be found in the literature (20, 21) and in the Materials and Methods section, respectively. The ionization paths leading to the same photoelectron energies, but leaving behind different cationic states, interfere and trigger coherent quantum motion of the remaining *N-1* electrons of the cation. The induced electronic dynamics depend on the involved eigenstates. With the pulse parameters used in the present study, most of the coherence produced in the glycine cation belongs to the ionic eigenstates in the 10a' band energy region (for details see Supplementary Materials), i.e. the states resulting from photoelectron emission from the 10a' orbital with binding energy of ~20 eV (the 10a' orbital is visualized in the center of Fig. 1B). Therefore, ionization results in an essentially-coherent superposition of a series of cationic eigenstates $\Psi_j$, each with ~5 - 30% contributions from the inner valence hole (1h) state and ~70 - 95% of a series of excited two-hole one-particle (2h1p) configurations (12). In the latter an additional bound valence electron is excited due to electron correlation. The time-dependent electronic wave packet describing the motion of the positive charge created in the photoionization step as a function of time *t* can be written in simple terms as a linear combination:

$$\Psi^{N-1}(t) = \sum c_j \, e^{-iE_j t} \, \Psi_j \qquad (1)$$

with expansion coefficients $c_j$. In this picture, the positive charge oscillates with periodicities $T$ given by the energy separation of the pairs of involved eigenstates, where $\hbar$ is Planck's constant over $2\pi$:

$$T = \frac{2\pi\hbar}{|E_i - E_j|} \qquad (2)$$

The *N-1* electron density, integrated over all electron coordinates but one, will also oscillate with the same periodicity due to constructive and destructive interference of the many-electron states contributing to the wave packet in Eq. (1).

The evolution of the positive charge in the molecular skeleton is followed in this work in two different measurements. On the one hand, we use x-ray absorption, where the transient electron hole states are probed, following a controlled time delay, by excitation of a strongly localized carbon 1s core electron into the 'empty' orbital via time-delayed fs x-ray pulses of the same color. As long as the electronic coherence, i.e. the electronic wave packet described by Eq. (1) is preserved, the probability of the resonant x-ray transition (C 1s → 10a', at 272.7 eV) that fills the inner-valence orbital and produces a carbon 1s core hole, depends on the time-dependent amplitudes of the cationic states with significant 1h-component localized on the carbon atom. The corresponding delay-dependent Auger yield is therefore a direct temporal fingerprint of the evolving many-electron wave packet in Eq. (1). Thus, the induced coherent electron dynamics following 10a' photoionization can be observed directly by recording the subsequent Auger electron emission as a function of x-ray pump – x-ray probe time delay. The basic experimental scheme flanked by *ab initio* simulations has been proposed by Cooper *et al.* (12) already in 2014. The theoretically predicted oscillation period $T$ is within (10 - 20) fs according to Refs. (12, 22, 23). On the other hand, we use non-resonant x-ray photoelectron spectroscopy, where the electron dynamics is monitored by photoelectron emission from sequential double photoionization processes. This turns out to be the dominant process (by a factor 10) in the time-resolved measurements at a central photon energy of 274.0 eV (off-resonance), although small contributions from the x-ray probe-induced Auger decay are present in this data set as well (see Fig. 1A and Fig. 1C).

Here we will report first the results from the x-ray absorption measurement and second the non-resonant sequential double ionization x-ray spectroscopy channel results. For the former a higher count rate was achieved, which enabled correlated multi-particle detection to isolate the channel of interest (see Materials and Methods). Otherwise the conditions in both of these independent measurements are the same.

In the x-ray absorption measurement based on the resonant transition followed by Auger decay probe, we set the FEL photon energy to the resonant x-ray probe transition (C 1s → 10a') at 272.7 eV and recorded delay scans over a range of 175 fs. In addition to the resonance effect (see Fig. 1C), this experiment benefited from a significantly higher density of glycine molecules in the focal volume (×100) compared to the experimental campaign probing the non-resonant channel. The increase was achieved by minimizing the distance between the molecular beam pipe and the FEL focus from ~10 mm to ~1 mm. This allowed us to collect 3 or more particle events per FEL pulse comprising the photoelectron at the kinetic energy corresponding to valence ionization of 10a' under investigation, the final $Gly^{2+}$ parent ion, and pump-probe events including Auger decay. Fig. 2 shows the result of this kinematically complete experiment spanning a 175-fs time scale. We counted the multi-particle events and plotted the relative change of the corresponding electron yield as a function of pump-probe delay in 1 fs steps. Oscillations of the detected electron signal in the time domain are clearly visible. The wavelet analysis of the electron signal shown in Fig. 2 reveals the frequency components that exist in the temporal domain (see Materials and Methods section for details). The short-time oscillation period was determined as $T = (19.6^{+1.5}_{-1.4})$ fs, in excellent agreement with the 20-fs oscillation period predicted by our *ab initio* many-electron calculations and exactly matches the period that was observed in the off-resonance sequential double ionization channel (see below). In addition, we observe that the 20 fs-period oscillations in the signal fade away after 40 - 50 fs and slower oscillations with a time period $T = (29.3^{+2.2}_{-2.0})$ fs become dominant and finally dissipate after 2 - 3 periods.

Fig. 3A shows the result on the shorter, 25 fs time scale probing the x-ray-induced electron dynamics off-resonance. Sequential double ionization is dominant (by a factor of 10 at least) over the Auger decay for the off-resonance probe FEL photon energy used in this part of our experiment, see Supplementary Materials Fig. S4. Here we counted the electrons with kinetic energies in four different ranges determined by the FEL spectral bandwidth: (i) from 259 to 263 eV, (ii) 252 to 256 eV, (iii) 235 to 239 eV and (iv) 231 to 235 eV and plotted the relative change of the electron yield in each energy bin (i-iv) as a function of pump-probe delay in 1 fs steps. Oscillations of the detected electron signal in the time domain are clearly visible. The curves for high-energy and low-energy electron emission show a $\pi$-phase shift with respect to each other, i.e. the electron yield in different energy ranges is maximized/minimized for different x-ray pump-probe delays as shown in Fig. 3B. In order to work out for which electron kinetic energy the phase jump occurs, the delay dependent electron spectra have been fitted in steps of 1 eV with sinusoidal functions and a 4-eV detection range keeping the oscillation period, amplitude and phase of the quantum beat as free parameters (see Materials and Methods section). The outcome is summarized in the Fig. 3C-F. We observe the $\pi$-phase jump at an electron kinetic energy of $(246\pm2)$ eV. Above and below this transition the relative change of the detected electron yield oscillates with a time period of $T = (19.6^{+2.2}_{-1.4})$ fs over the full kinetic energy range of electron emission from 224 to 264 eV shown in Fig. 1A. This time period is in close agreement to the early time oscillations measured in the independent x-ray absorption measurement presented in Fig. 2.

We have simulated the experimental observables using the measured FLASH pulse parameters and including the various ionization channels. The numerical *ab initio* results on the probe-induced electron-yield modulation in pump-ionized glycine molecules as a function of x-ray pump-probe delay are summarized in the Fig. 4A-C. Our theoretical simulations show that the phase jump observed in the off-resonance experiment is a property of the sequential double ionization, see further details in the Supplementary Materials section 2.2. Both channels monitor the same beating frequency of the cationic eigenstates.

The calculations were performed using the time-dependent B-spline restricted correlation space (RCS) – algebraic diagrammatic construction (ADC) simulation method (24-26), which provided the populations, degrees of coherence and relative phases between each pair of accessible cationic states computed using the ADC(2,2) method of ref. (27), as well as the relative change of electron kinetic energy spectra as a function of pump-probe time delay as recorded in the measurements. The simulation combines the accurate description of the photoelectron continuum by means of the B-spline basis set (28) and the ADC description of electron correlation, but in view of the required computational effort is restricted to a single (equilibrium) geometry of the Gly I conformer. The lack of geometry averaging is probably the main source of discrepancy between the measured (Fig. 3) and simulated (Fig. 4) probe electron spectra. The agreement between both on-resonance and off-resonance experiments and the theory in terms of the initial period of the oscillations of the many-electron wave packet allows us to interpret the initial 25 fs of the quantum evolution of the glycine cation as resulting from the electronic coherence. The amplitude of the observed oscillation reflects the contribution of the 10a' photoionization channel and the magnitude of the electronic coherence produced within the 10a' band of many-electron states to the overall signal, see further details in section S2.2 and Fig. S3 in the Supplementary Materials section.

We note that within the time-dependent B-spline ADC(2)x scheme used for simulating the probe step of the experiment, the absolute binding energies of the final dicationic eigenstates are not reproduced with high accuracy and may easily be off by up to ~10 eV [this is in contrast to the accurate cationic eigenenergies obtained using the ADC(2,2) method and used in the simulations of the pump step]. Therefore, we plot the simulated time-dependent electron spectra presented in Fig. 4B and Fig. 4C relative to the theoretically derived phase jump energy (~255 eV), which is above the observed experimental value of (246±2) eV. Furthermore, the ionization dynamics for pump-probe time delays of a few fs, where the x-ray pulses strongly overlap, cannot be accurately described in our model. The reason is that theoretically the pump and probe simulations are carried out independently and the probe simulation assumes that the pump-induced ionization is completed. This is also the reason why the onset of the observed electron-yield oscillation slightly differs between experiment and theory. We also note that according to theory the Auger spectra, as well as the sequential double ionization spectra do not significantly depend on whether the inner-valence hole is located on the α-carbon or on the carboxyl carbon (see Supplementary Materials). Although the resonant 4a' → 9a' transition energy for inner-shell excitation of the carboxyl group is close to the FEL photon energy and therefore considered in the simulation, the FEL spectral bandwidth of less than 1.2 eV (FWHM) is too small to form a coherent superposition of the two C 1s core orbitals 4a' and 5a' with high efficiency. Their relative chemical shift is on the order of ~2.9 eV, which gives a time period of ~1.4 fs for the corresponding quantum beat that cannot be resolved in the present experiment. Nevertheless, the contribution from adjacent 9a' configurations to the observed dynamics is fully taken into account in our simulation and is also visible in static experiments, where two electrons are measured in coincidence (Fig. 5 in the Materials and Methods section).

Our further simulations take into account the nuclear geometry (but not nuclear dynamics) effects on the temporal evolution of the many-electron wave packet. First, we note that at an average oven temperature of 160 °C in our gas phase experiment only two conformers commonly referred to as Gly I and Gly III are expected to contribute to the molecular response at a ratio of approximately 2:1 (12, 29). Theoretical investigations have shown that the ionization out of orbital 10a' gives rise to different many-electron dynamics in Gly III and Gly I with time scales varying from about 10 to 20 fs, respectively (23). Therefore, the time-resolved electron spectroscopic data presented here gives the effective oscillation period $T$ for the mix of conformers present in the molecular beam. In accordance, we have also calculated the survival probability of the pump-prepared ionic density matrix performing a 2:1 weighted average over the Gly I and Gly III conformers. Furthermore, our simulation of the survival probability also included the effect of the quantum uncertainty in the nuclear geometry due to zero-point energy. To this end, we have used the normal modes of Gly I and Gly III conformers obtained using the Molpro quantum chemistry software package (30). Our simulations show that the oscillations of the many-electron wave packet survival probability survive both conformer and the geometry uncertainty averaging for over at least 25 fs, in agreement with the experimental results, see Fig. S5 in the Supplementary Materials. It should be noted that in some other systems, the effect of the quantum uncertainty in the nuclear positions led to a much faster dephasing of the coherent oscillations of the many-electron wave packets. For example, in isopropanol (31), it has been found that while the initial quasi-exponential decay of the inner-valence electron hole is barely affected by the averaging over nuclear geometries, the subsequent quantum revivals of the hole population predicted theoretically at the equilibrium geometry of the neutral molecule are completely destroyed by the averaging over an ensemble of geometries representing a finite-uncertainty vibrational state. Similarly in paraxylene (11), averaging over the nuclear geometries has been shown to lead to the very efficient dephasing of the hole oscillations due to coherent population of two electronic states.

**Discussion**

This work contributes to the solution of the controversy on the possible time scale of the coherent many-electron dynamics leading to charge migration in ionized polyatomic molecules. Whereas previously such dynamics, spanning 10s of femtoseconds, was only detected indirectly, e.g. via nuclear rather than electronic observables (3), here we present results of two direct modes of detection both based on the electronic observables, see Fig. 2 and Fig. 3. We have shown that the electron-yield oscillations monitoring the quantum wave packet dynamics of the positive charge in the photoionized glycine cation persist for at least 40 fs. Our work, therefore, puts the idea of the long-time coherent electronic molecular dynamics triggered by photoionization on a much firmer ground than it was previously possible. It should be noted another study of charge migration in a completely different ("frustrated Auger") regime in isopropanol was recently published (31) that also uses the spectroscopic technique proposed theoretically in ref. (12) to observe highly transient hole states. However, the physical regime which ref. (31) addresses is the short-time one, covering only the first few fs of the coherent evolution.

The presented observations of the oscillations in the photo- and Auger electron yields prove the existence of coherent dynamics in photoionized glycine extending over tens of fs. To extract the exact nature of the coherence leading to these observed oscillations in an experiment probing only the electronic degrees of freedom we require theoretical analysis. Our high level ADC(2,2) treatment is essential to capture with sufficient accuracy the electronic dynamics, as it provides the proper treatment of the 1h and 2h1p components of the 10a' state essential for this particularly challenging inner-valence spectral region characterized by partial breakdown of the molecular orbital picture of ionization. A full quantum mechanical simulation of vibronic coherence and vibronic dynamics following glycine photoionization, coupled to this accurate electronic calculation, is presently out of reach. Therefore, we limited the treatment of the nuclear degrees of freedom in our theoretical investigation to averaging over the zero-point energy spread in the nuclear positions, which has been previously shown to lead to dephasing of the electronic coherence (11). Our *ab initio* theoretical investigation showed that the period of the observed initial oscillations (~20 fs) is fully consistent with the Born-Oppenheimer picture of an electronic process taking place over a distribution of nuclear geometries.

While the oscillation period observed in the electronic signals is also similar to that of the glycine C=O stretch, it cannot be explained by a coherent excitation of vibrations in the cation on a potential energy surface of a single electronic state, because coherent population of several bands of electronic states has been firmly established by the pump simulation (see Fig. S1 in the Supplementary Materials). At the same time, it would be naive to attribute the observed coherence to purely electronic one as modelled here within the Born-Oppenheimer picture. Indeed, energy spacings in the energy region of partial breakdown of the molecular orbital picture to which the 10a' states belong are of the same order of magnitude as some of the vibrational quanta and the two degrees of freedom are expected to strongly couple, resulting in quantum eigenstates represented by linear superpositions of electronic states dressed by vibrational excitations. Non-adiabatic dynamics and its effect on the vibronic coherence has been studied by Mukamel and co-workers in the case of core ionization or excitation (32-35), where breakdown of the molecular orbital picture is not typical. In our work, we probe experimentally, via electronic degrees of freedom, the cationic state coherence resulting from the inner-valence ionization. Our many-electron simulations show that this coherent oscillation period of the probe electron signal can be successfully approximated by the coherent dynamics of the Born-Oppenheimer electronic states averaged over the zero-point energy distribution of the nuclear geometries. However, the full coherence brought to life by the few-fs ionization of glycine in the inner valence region is, in all probability, of mixed electronic and vibrational character. Full characterization of this coherence theoretically within the highly computationally demanding inner-valence energy region is currently beyond reach and should be subject of future theoretical studies. One can, nevertheless, speculate on the basis of the experimental results presented here, that the long time span of the coherent signal oscillations observed in this work may well be a result of the similarity in the time scales of the electronic and vibrational dynamics.

## Materials and Methods

**Experimental design** The experiment was carried out at the FL24 beamline of the FLASH2 free-electron laser (FEL) at DESY in Hamburg (18). The FEL is operated in 10 Hz bursts with trains of 400 electron bunches each with 1 μs separation, in a special short-pulse mode with low electron bunch charge of a few tens of pC. Each bunch generates a coherent FEL pulse with sub-5 fs duration at 274.0 eV (off-resonance) and 272.7 eV (on-resonance) photon energy and an average pulse energy of about 5 μJ (4000 photon pulses per second in total). The FEL wavelength is tuned by changing the gap between the magnetic poles of the FEL undulator magnets, while the electron beam energy is kept constant at 1.24 GeV. The scheme of the experimental setup is shown in Fig. 6.

We used a split-and-delay unit consisting of two interleaved lamellar mirrors in the unfocused beam to generate two replicas of collinearly propagating FEL pulses with a controllable delay (17). The resulting pump-probe sequences were focused into the interaction region by a toroidal mirror with a focal length of about 5.7 m. Both, the lamellar-mirror pair and the focusing mirror were nickel-coated and operated at 8° angle of incidence to the surface.

For the experiments, >98.5% glycine was acquired from Sigma-Aldrich and used without further purification. The glycine molecules were delivered to the interaction region by evaporation of glycine powder at 160 °C in a resistively heated oven. The resulting vapor pressure of glycine in the $10^{-3}$ mbar range produced an effusive beam of sufficient density.

The photoelectrons and Auger electrons generated in the interaction region were detected by a microchannel plate (MCP) detector in a magnetic-bottle type time-of-flight (TOF) electron spectrometer. It allows the application of retardation fields to provide the necessary resolution for fast electrons with several hundred eV kinetic energy. The MCP signal was digitized by a fast analog-to-digital converter and stored in the FLASH data acquisition system. The electron kinetic energies were calibrated with the well-known photoelectron spectrum of argon. The setup also allowed for simultaneous measurement of TOF mass spectra, which were calibrated by means of mass spectrometry using different rare gases.

To generate the shortest possible pulses FLASH uses a separate photoinjector laser optimized for low bunch-charge operation. The low charge enables a very high bunch compression due to smaller space-charge effects. The short electron bunch reduces the number of individual longitudinal modes present in the FEL pulses, in the ideal case down to a single mode. This is referred to as single-spike SASE or single-mode SASE operation. It increases the longitudinal coherence as the slippage between the electrons and the light in the undulator is larger than the bunch length so the light emitted from the tail of the electron bunch reaches the head of the bunch within a few gain lengths and clearly before saturation. Single-spike operation delivers sub-5 fs pulses with a high degree of longitudinal coherence from a SASE source as the whole electron bunch interacts with the light from the same mode (20, 21).

During our beamtime we used the Online Photoionization Spectrometer (OPIS) system in the FLASH2 tunnel for photon diagnostics (18). OPIS consists of four electron time-of-flight spectrometers arranged in a cross shape to monitor the FEL wavelength and bandwidth by detecting photoelectrons from a thin rare gas target. In order to evaluate the FEL spectral bandwidth we correlated events in which two electrons were detected in opposite lying TOF spectrometers. Each photoelectron was assigned to either the Argon $2p_{1/2}$ or $2p_{3/2}$ orbitals and only matching pairs originating from the same orbital of two atoms within one FEL shot were considered. For the dataset shown in Fig. 3 it results in a total number of $\sim 2.6 \times 10^7$ electron pairs, which are plotted in Fig. 7. For these pairs, the differences of their respective photoelectron kinetic energies were calculated and are shown as a histogram. The width of the equidistant energy bins was chosen to approximate the resolution of the analog-to-digital converter with a sampling rate of 7 GHz. This procedure results in an autocorrelation of the FEL shots in the spectral domain. Assuming a Gaussian distribution, the FEL spectral bandwidth during the data acquisition in the glycine experiment can be calculated to be 1.24 eV (FWHM) or about 0.45% of the central photon energy tuned in the range between 269 eV and 281 eV. We note that this number still includes the instrument functions of the OPIS spectrometers as well as a broadening by the finite size beam profile and pointing jitter of the FEL and is therefore an upper limit.

**Data analysis and fitting** In Fig. 1A each spectrum was normalized by the respective number of detected electrons to account for any fluctuations in FEL pulse energy and other FEL parameters. The single spectra for each wavelength were measured for about 5 to 6 minutes. The average FEL wavelength per bunch train drifted less than 0.1% around the central wavelengths. Each spectrum shown in Fig. 1A is stacked vertically with its offset corresponding to the photon energy set by the undulator gap.

Instead of using the Jacobi transformation for the electron TOF to kinetic energy conversion, the equidistant TOF bins were resampled to equidistant energy bins with a linear filter to avoid aliasing effects. The energy bin size of 0.3 eV was chosen to match the energy width of the first time-of-flight bin in the spectra shown in Fig. 1A.

In Fig. 3A the electron spectra were divided by the total number of detected electrons for each delay to account for any FEL fluctuations and afterwards the residual gas contribution was subtracted. Here, the measurement time was 10 minutes per delay. The y-axis shows the relative change of the electron yield compared to the average yield in percent. All four oscillations were fitted with a sinusoidal curve of the form:

$$f_i(\Delta t) = a_i \cdot \sin(\omega_i \cdot \Delta t + \varphi_i)\,, \tag{3}$$

where $a_i$ is the amplitude, $\varphi_i$ a phase offset, $\omega_i = \frac{2\pi}{T_i}$ and $T_i$ the period of the oscillation for $i \in [1, 4]$. As the average spectrum over all delays was subtracted for every single delay the resulting relative change of the electron yield in percent naturally oscillates around 0.

The oscillation amplitudes, periods, phases and the goodness $R_i^2$ of the fits are summarized in Table 1.

For Fig. 3C the kinetic energy range of 224 to 264 eV was split into overlapping intervals with 4-eV width in steps of 1 eV. Each of these delay dependent electron spectra was fitted according to the formula used for Fig. 3A. The result of the 37 individual fits is plotted in Fig. 3C with the central energy of the 4-eV energy range on the y-axis.

All of the kinetic energy ranges from Fig. 3C were also fitted with sinusoidal curves of the same period but free phase and amplitudes according to:

$$f_i(\Delta t) = a_i \cdot \sin(\omega \cdot \Delta t + \varphi_i), \qquad (4)$$

where $a_i$ is the amplitude and $\varphi_i$ a phase offset for $i \in [1, 37]$, $\omega = \frac{2\pi}{T}$ and $T$ the period of the oscillation.

The resulting period $T$ of this fit is indicated in Fig. 3A and Fig. 3E as black dashed lines. The shown error values are the mean of the upper and lower 95% for the 37 individual sinusoidal fits of overlapping energy intervals.

The fitted period presented in Fig. 3 within two standard deviations, $T = (19.6^{+2.2}_{-1.4})$ fs, corresponds to an energy level separation of the two cationic eigenstates $\Psi_i$ and $\Psi_j$ of $\Delta E \approx (0.21 \pm 0.02)$ eV according to equation (2).

To identify the somewhat weaker probe-induced Auger channel at the central photon energy of 274.0 eV (off-resonance) experimentally demonstrating the element specificity and orbital selectivity of the 1s $\rightarrow$ 1h resonant transition, we selected FEL shots in which two electrons were detected in coincidence. Out of ~$6 \times 10^5$ electrons created in the interaction of glycine molecules with ~$1 \times 10^8$ FEL pulse pairs we identified ~7600 coincidences integrated over all time delays. In Fig. 5 the kinetic energy of the first (faster) and second (slower) electron are plotted on the x- and y-axis, respectively. Among the many possible multielectron excitation pathways and decay channels creating doubly charged final states, particularly those cascades starting with photoionization of the 9a' and 10a' molecular orbitals are significantly enhanced in the coincidence map as indicated in Fig. 5. The signature of a pump-probe event including Auger decay is the observation of a photoelectron at kinetic energy corresponding to valence ionization of 10a' (or 9a') in combination with an Auger electron that is downshifted in energy by the difference of the 1st and 2nd ionization potential.

The coincidence event map is obtained as follows: Every single-shot kinetic energy spectrum with two electrons is represented by a row vector $S(E_1, \dots, E_N)$, where $1 \dots N$ are the indices of energy bins. It is transformed into an $N \times N$ matrix $M = S^T S$. Then all these single-shot matrices, which are symmetric with respect to the diagonal, are summed up and the diagonal line together with the upper triangle are removed for clarity. Columns and rows of the resulting matrix represent the energies of the first electron and the second electron, respectively. The values of elements $p_{mn}$ of the matrix contain the number of double events, where the first electron had the kinetic energy $E_n$ and the second electron had the kinetic energy $E_m$.

Based on the knowledge gained from the coincidence map of the static experiment discussed above we searched our data recorded at the FEL photon energy of 272.7 eV (on-resonance) for two or more electron detection simultaneously to the appearance of the doubly-charged glycine parent ion $(Gly^{2+})$ per FEL shot. Then, we filtered these triple events fixing the kinetic energy of one electron to the valence ionization of 10a' within a detection window of 4 eV. The resulting electron yield is divided by the total number of detected electrons for each delay to account for any FEL fluctuations. Afterwards the difference between this signal and the average over the full delay scan is plotted in order to derive the relative changes of the electron yield as a function of x-ray pump x-ray probe delay in 1-fs steps, which is shown in Fig. 2. Continuous wavelet transform (CWT) allows to study the magnitude evolution of a nonstationary signal at scaling frequencies. We employ generalized Morse wavelets as described by Lilly and Olhede (36) with symmetry parameter $\gamma = 3$ and time-bandwidth product $P^2 = 60$. The CWT uses 48 voices per octave. The 'cone of influence' (COI), where part of the wavelet (in time domain) extends past the finite signal trace, is given as a darker shaded area. The boundaries of the COI are chosen as the points, where the autocorrelation magnitude of the respective wavelet decays by $1/e$. Peaks in the shaded area are likely underestimated.

M. Braune, A. Brinkmann, O. Brovko, T. Bruns, P. Castro, J. Chen, M. K. Czwalinna, H. Damker, W. Decking, M. Degenhardt, A. Delfs, T. Delfs, H. Deng, M. Dressel, H.-T. Duhme, S. Düsterer, H. Eckoldt, A. Eislage, M. Felber, J. Feldhaus, P. Gessler, M. Gibau, N. Golubeva, T. Golz, J. Gonschior, A. Grebentsov, M. Grecki, C. Grün, S. Grunewald, K. Hacker, L. Hänisch, A. Hage, T. Hans, E. Hass, A. Hauberg, O. Hensler, M. Hesse, K. Heuck, A. Hidvegi, M. Holz, K. Honkavaara, H. Höppner, A. Ignatenko, J. Jäger, U. Jastrow, R. Kammering, S. Karstensen, A. Kaukher, H. Kay, B. Keil, K. Klose, V. Kocharyan, M. Köpke, M. Körfer, W. Kook, B. Krause, O. Krebs, S. Kreis, F. Krivan, J. Kuhlmann, M. Kuhlmann, G. Kube, T. Laarmann, C. Lechner, S. Lederer, A. Leuschner, D. Liebertz, J. Liebing, A. Liedtke, L. Lilje, T. Limberg, D. Lipka, B. Liu, B. Lorbeer, K. Ludwig, H. Mahn, G. Marinkovic, C. Martens, F. Marutzky, M. Maslocv, D. Meissner, N. Mildner, V. Miltchev, S. Molnar, D. Mross, F. Müller, R. Neumann, P. Neumann, D. Nölle, F. Obier, M. Pelzer, H.-B. Peters, K. Petersen, A. Petrosyan, G. Petrosyan, L. Petrosyan, V. Petrosyan, A. Petrov, S. Pfeiffer, A. Piotrowski, Z. Pisarov, T. Plath, P. Pototzki, M. J. Prandolini, J. Prenting, G. Priebe, B. Racky, T. Ramm, K. Rehlich, R. Riedel, M. Roggli, M. Röhling, J. Rönsch-Schulenburg, J. Rossbach, V. Rybnikov, J. Schäfer, J. Schaffran, H. Schlarb, G. Schlesselmann, M. Schlösser, P. Schmid, C. Schmidt, F. Schmidt-Föhre, M. Schmitz, E. Schneidmiller, A. Schöps, M. Scholz, S. Schreiber, K. Schütt, U. Schütz, H. Schulte-Schrepping, M. Schulz, A. Shabunov, P. Smirnov, E. Sombrowski, A. Sorokin, B. Sparr, J. Spengler, M. Staack, M. Stadler, C. Stechmann, B. Steffen, N. Stojanovic, V. Sychev, E. Syresin, T. Tanikawa, F. Tavella, N. Tesch, K. Tiedtke, M. Tischer, R. Treusch, S. Tripathi, P. Vagin, P. Vetrov, S. Vilcins, M. Vogt, A. de Zubiaurre Wagner, T. Wamsat, H. Weddig, G. Weichert, H. Weigelt, N. Wentowski, C. Wiebers, T. Wilksen, A. Willner, K. Wittenburg, T. Wohlenberg, J.

## Acknowledgements

**General**: We thank the late Prof. Dr. Wilfried Wurth for many inspiring discussions.

**Funding:** This work was funded by the Deutsche Forschungsgemeinschaft (DFG, German Research Foundation) through the Cluster of Excellence 'Advanced Imaging of Matter' (EXC 2056 - project ID 390715994), the collaborative research center 'Light-induced Dynamics and Control of Correlated Quantum Systems' (SFB-925 – project 170620586), the projects KI 482/20-1 and LA 1431/5-1, and by the Federal Ministry of Education and Research of Germany under Contract No. 05K10CHB.

**Author contributions:** T.L. led the project. D.S., S.U. and A.P. prepared the endstation. S.Dei. and J.V. developed the magnetic-bottle electron spectrometer. J.R.S. and M.K.C. prepared the FEL. M.K., S.Due, S.T., S.P., I.B., M.M. and M.B. prepared the photon beamline and ran the online photon diagnostics. S.U., A.P., C.H. and D.K. developed the split-and-delay unit coated with nickel by L.B. and R.R. The experiment was performed by D.S., S.U., A.P., K.M., S.Dei., A.H., G.H., P.S. and T.L. The data were analyzed by D.S and S.U. with support from P.S., G.H. and A.K. The results were interpreted by D.S., S.U., A.P. and T.L. Theory and modelling were performed by M.R. and P.K. with support from V.A. and J.P.M. The corresponding sections in the supplementary materials were written by M.R. The manuscript was written by T.L and V.A. with contributions from J.P.M., D.S., S.U., and A.P., as well as input from all authors.

**Competing interests:** The authors declare that they have no competing interests.

**Data and materials availability:** The authors declare that the main data supporting the findings of this study are available within the article and its supplementary information file. Extra data are available from the corresponding author upon reasonable request.

## Figures and Tables

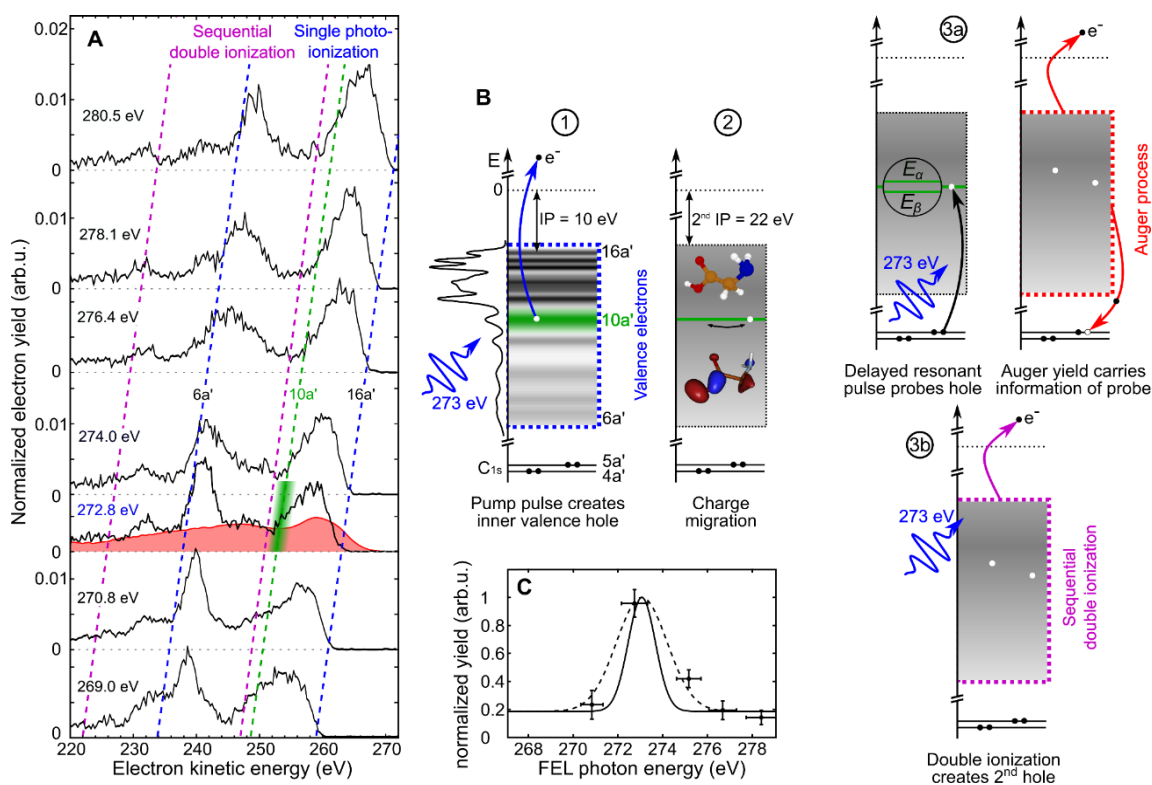

**Fig. 1. X-ray photoelectron spectroscopy of glycine. (A)** Kinetic energy distributions of detected electrons as a function of FEL photon energy using single pulses, i.e. $\Delta t = 0$. Tilted lines (dashed blue, dashed violet) mark the energy ranges of single photoionization and sequential double photoionization of valence and inner-valence electron orbitals (6a' - 16a', 1a'' - 4a''), respectively. An experimental Auger electron spectrum (red) for C 1s core vacancies (4a', 5a') is reproduced from ref. (15). The overlap of the bandwidth of the ionizing radiation with the core-to-10a' valence transition is indicated (green). **(B)** A single-color pump-probe scheme is applied to track charge dynamics initiated by photoelectron emission from the 10a' molecular orbital. The total cascade involves several different processes: (1) photoionization, (2) charge migration triggered by coherent population of ionized states during photoionization, (3a) resonant carbon 1s core-hole excitation and Auger decay, as well as (3b) sequential double photoionization of valence electrons. The individual steps are depicted as a sequence of energy diagrams. The density of states is indicated by using the experimental data reproduced from ref. (14). **(C)** Two-electron coincidences plotted as a function of FEL photon energy with one electron being detected at the kinetic energy corresponding to valence ionization of 10a'. These data are recorded in a second set of measurements with significantly increased data acquisition time by a factor ~100 compared to (A). The resonance is fitted with a Gaussian envelope (dashed line) and deconvoluted (solid line) with respect to the spectral bandwidth of the FEL pulses (see Fig. 7 in Materials and Methods). The data are normalized to the FEL pulse energy to account for FEL fluctuations as well as to the maximum signal on resonance.

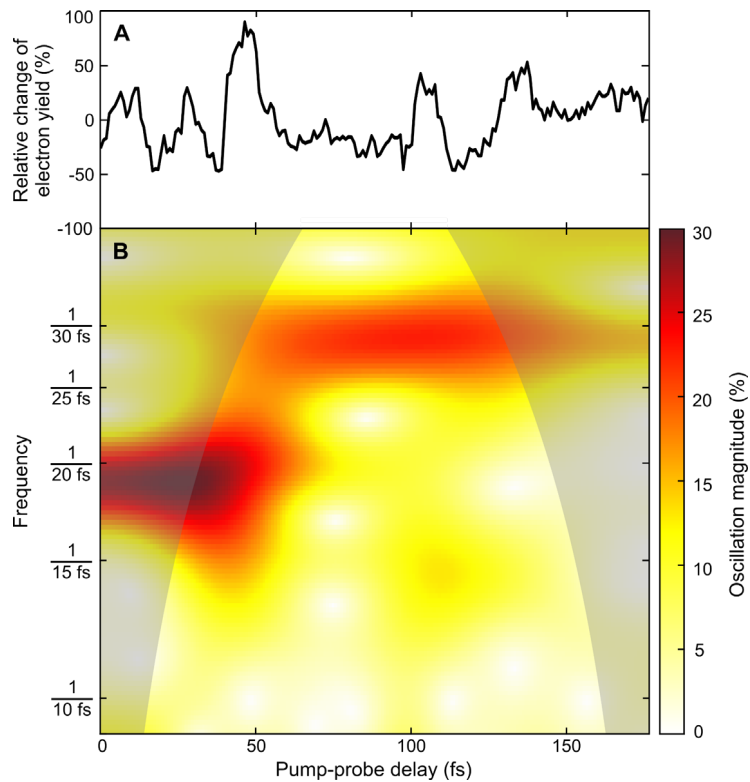

**Fig. 2. Electron signal oscillations induced and monitored with 272.7 eV photons and wavelet analysis. (A)** Relative change of the detected electron yield correlated with the generation of a $Gly^{2+}$ parent ion and recorded together with an electron at the kinetic energy corresponding to valence ionization of 10a' as a function of x-ray pump-probe delay in 1-fs steps (black line). **(B)** The continuous wavelet transform (see Materials and Methods) monitors the magnitude evolution of the nonstationary signal at scaling frequencies (false color plot). The 'cone of influence', where part of the wavelet in time domain extends past the finite signal trace, is given as a darker shaded area. Peak structures in the shaded area of the color plot are likely underestimated. Two dominant time periods of $T = (19.6^{+1.5}_{-1.4})$ fs (over the first 40 fs) and $T = (29.3^{+2.2}_{-2.0})$ fs (later on) can be extracted, with the initial period being in excellent agreement with the *ab initio* theory explaining the oscillations as a result of electronic coherence averaged over the nuclear geometries (for details, see the text).

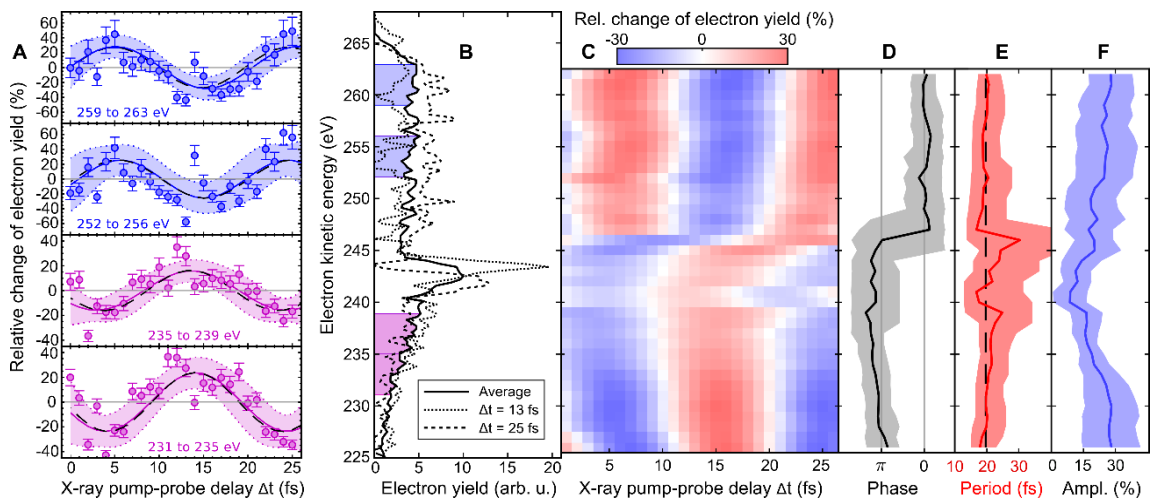

**Fig. 3. Phase-sensitive electron wave packet interferometry of glycine with 274.0 eV photons. (A)** Relative change of the detected electron yields in kinetic energy ranges with contributions from Auger emission superimposed with sequential double ionization as a function of x-ray pump-probe delay. The solid lines are the sinusoidal fitting functions of time period $T$. The 95% confidence bounds are given by dotted lines. **(B)** Kinetic energy distributions of detected electrons recorded for pump-probe delays $\Delta t = 13$ fs and $\Delta t = 25$ fs compared with the average electron spectrum over all delays. **(C)** Delay dependent electron spectra fitted in steps of 1 eV with sinusoidal functions and 4 eV detection range having the oscillation period, amplitude and phase as free parameters. The fitting functions are presented as a false color plot. **(D)** Phases, **(E)** periods and **(F)** amplitudes of the sinusoidal functions (solid black, solid red and solid blue line) depending on the kinetic energy of detected electrons. The 95% confidence bounds are given by the shaded areas. The black dashed lines shown in (A) and (E) are the result of using a single time period $T = (19.6^{+2.2}_{-1.4})$ fs for fitting the complete data set.

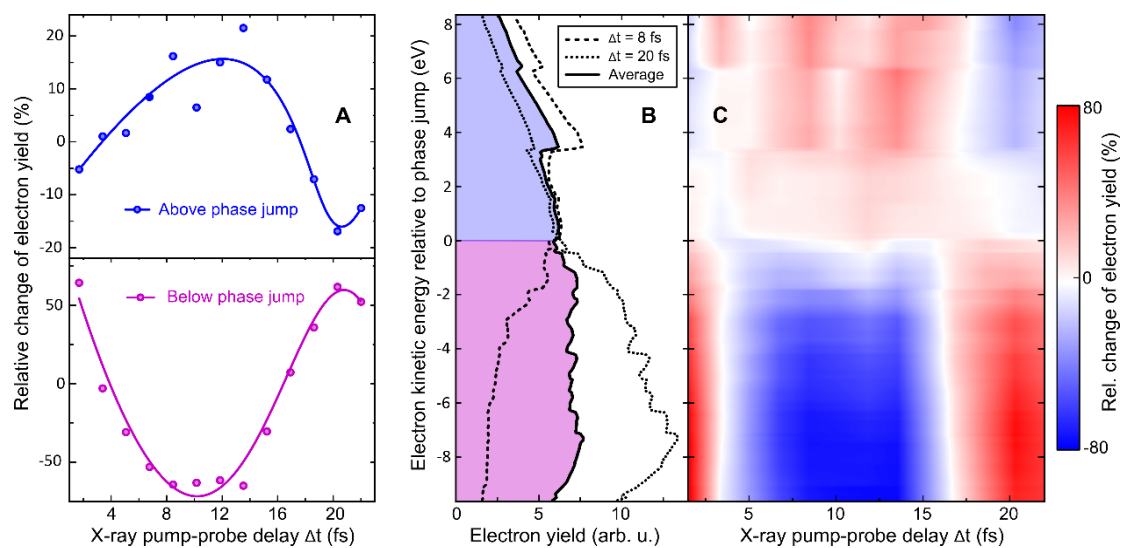

**Fig. 4. *Ab-initio* TD B-spline RCS-ADC simulation of experimental observables using the measured FLASH pulse parameters in the off-resonance probe scenario. (A)** Relative change of the detected electron yields in kinetic energy ranges above and below the phase jump including contributions from Auger emission and sequential double ionization as a function of x-ray pump-probe delay. The solid lines are guide-to-the eye showing the onset of an oscillatory behavior with a time period of ~20 fs. **(B)** Kinetic energy distributions of emitted electrons simulated for pump-probe delays $\Delta t = 8$ fs and $\Delta t = 20$ fs compared with the average electron spectrum over all calculated delays. **(C)** Delay dependent electron spectra presented as a false color plot system (for details see Supplementary Materials).

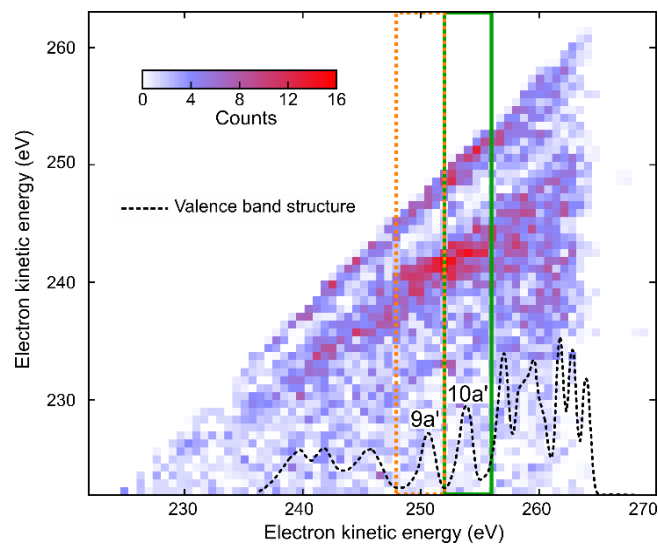

**Fig. 5. Two-electron coincidence spectroscopy of glycine with 274.0 eV photons.** Coincidence map of two-electron detection. Photoelectron (10a', 9a') – Auger electron coincidences following the resonant x-ray transition (1s → 1h) are indicated (green and orange squares). The density of valence states is indicated (dotted black) by using the experimental data reproduced from ref. (14). Valence-electron emission from x-ray probe-induced sequential double ionization contributes to an uncorrelated background with respect to the (initial) x-ray pump-induced photoionization events, which allows observing the fingerprint of the probe-induced Auger decay. The map also shows an enhanced probability for the detection of two (photo)electrons with an equal kinetic energy, which are the likely result of a false coincidence between two valence-ionization events.

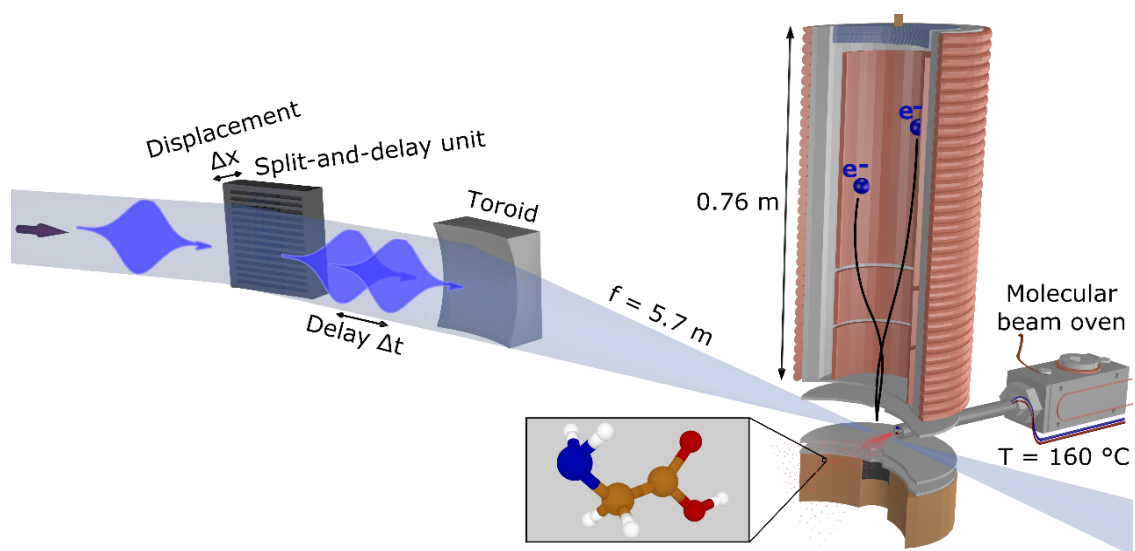

**Fig. 6. Schematic setup.** Experimental setup for the single-color, femtosecond x-ray pump-probe measurements of glycine at FLASH2.

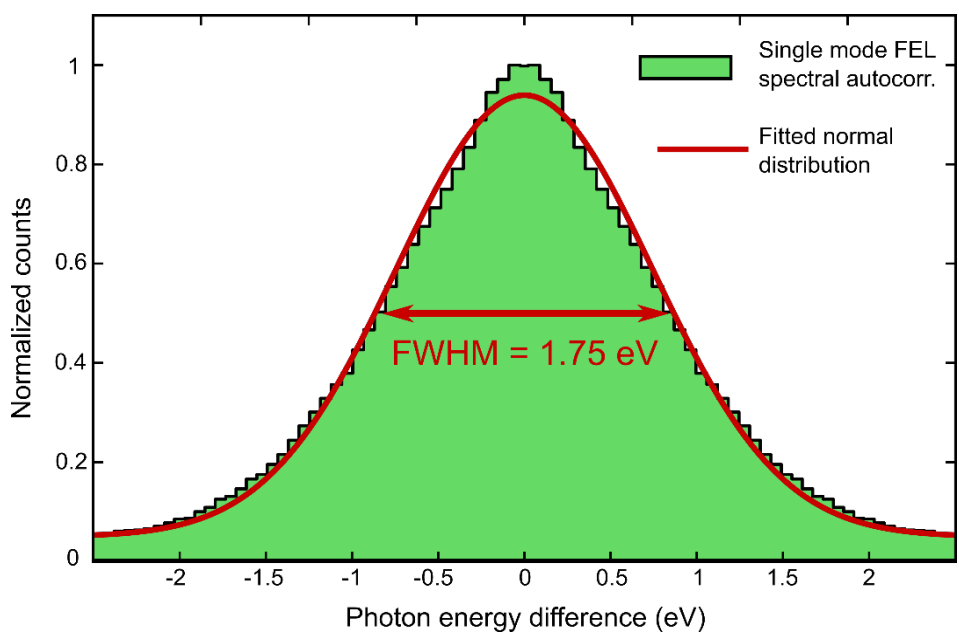

**Fig. 7. FEL photon energy autocorrelation** for the dataset shown in Fig. 3. Photoionization from the Argon $2p_{1/2}$ and $2p_{3/2}$ orbitals was used to infer the FEL photon energy from four TOF spectrometers.

**Table 1. Fitting parameters of sinusoidal curves of the form of equation (3).** The oscillation amplitudes, periods, phases and the goodness $R_i^2$ of the fits describing the experimental data shown in Fig. 3A are summarized.

| Energy range | Amplitude | Period | Phase | Goodness |
|---|---|---|---|---|
| 259 to 263 eV | $a_1 \approx (28 \pm 11)\%$ | $T_1 \approx (20.6^{+3.8}_{-2.8})$ fs | $\varphi_1 \approx \quad (0.0 \pm 0.2)\pi$ | $R_1^2 \approx 0.58$ |
| 252 to 256 eV | $a_2 \approx (26 \pm 14)\%$ | $T_2 \approx (18.8^{+5.4}_{-3.4})$ fs | $\varphi_2 \approx (-0.1 \pm 0.3)\pi$ | $R_2^2 \approx 0.37$ |
| 235 to 239 eV | $a_3 \approx (16 \pm \ 6)\%$ | $T_3 \approx (21.4^{+6.2}_{-3.9})$ fs | $\varphi_3 \approx \quad (1.2 \pm 0.3)\pi$ | $R_3^2 \approx 0.55$ |
| 231 to 235 eV | $a_4 \approx (24 \pm \ 9)\%$ | $T_4 \approx (20.5^{+4.6}_{-3.2})$ fs | $\varphi_4 \approx \quad (1.1 \pm 0.3)\pi$ | $R_4^2 \approx 0.58$ |

Supplementary Materials for

# Electronic Quantum Coherence in Glycine Molecules Probed with Ultrashort X-ray Pulses in Real Time

David Schwickert, Marco Ruberti*, Přemysl Kolorenč, Sergey Usenko, Andreas Przystawik, Karolin Baev, Ivan Baev, Markus Braune, Lars Bocklage, Marie Kristin Czwalinna, Sascha Deinert, Stefan Düsterer, Andreas Hans, Gregor Hartmann, Christian Haunhorst, Marion Kuhlmann, Steffen Palutke, Ralf Röhlsberger, Juliane Rönsch-Schulenburg, Philipp Schmidt, Sven Toleikis, Jens Viefhaus, Michael Martins, André Knie, Detlef Kip, Vitali Averbukh, Jon P. Marangos, Tim Laarmann*

*Corresponding author. Email: m.ruberti11@imperial.ac.uk; tim.laarmann@desy.de

**This PDF file includes:**

Supplementary Text
Figs. S1 to S5

**Supplementary Text**

**S1. Theoretical methods.**

Advanced theoretical modelling and simulation of the photoionization dynamics of the glycine molecule triggered by the x-ray pump and probe laser pulses is performed here, from first-principles, within the framework of the cutting-edge time-dependent B-spline restricted correlation space (RCS) – algebraic diagrammatic construction (ADC) ab initio methods (24-26). Both interactions of the neutral glycine molecule and its cation with the pump and probe x-ray femtosecond pulses, respectively, are described from first-principles. The simulations of photoionization during the pump and probe steps are carried out, within the frozen-nuclei approximation, at the equilibrium geometry of the Gly I conformer. We also analyze the effect of the geometry-spread of the ground state nuclear wavefunction on the survival of the quantum electronic coherences in the 10a' and 9a' ionized bands (see Section S3).

**S1.1 First ionization of the neutral species by the x-ray pump pulse.**

Within B-spline RCS-ADC, the single-electron orbitals are expanded in a multicenter B-spline basis set (24, 28), designed to accurately describe both bound and ionized states of the neutral molecule. The resulting Hilbert space of Hartree-Fock (HF) virtual orbitals ($\phi_\alpha$), is subsequently partitioned into two orthonormal subspaces: the restricted correlation space (RCS) ($\chi_\alpha$), designed to accurately describe the localized short-range correlation of the system, and its orthonormal complement, the ionization space (IS) $\left(\psi_\mu\right)$, which describes the long-range part of the photoelectron wavefunction and complements the RCS in order to represent the electronic wavefunction over the entire spatial region (24). Here and in what follows, the $\alpha, \beta, \ldots$ and $\mu, \nu, \ldots$ indices refer to the virtual space (unoccupied) RCS and IS orbitals, respectively, whereas i, j, k, ... , indicate the occupied HF molecular orbitals.

Within the time-dependent (TD) B-spline RCS-ADC(n) approach to molecular photoionization dynamics (25, 26), the 3D time-dependent Schrödinger equation (TDSE) for an N-electron polyatomic molecule interacting with an ultrashort laser pulse

$$i\, \hbar\, \frac{\partial \mid \Psi^N(t) \rangle}{\partial t} \;=\; \hat{H}^N_{Pump}\,(t)\; \mid \Psi^N\,(t)\,\rangle \qquad\qquad (1)$$

is solved by expanding the TD many-electron wavefunction in the basis of the ground and excited RCS-ADC(n) states (25, 26)

$$| \Psi^N(t) \rangle = \sum_{m,\mu} c_{m,\mu}(t) \, \hat{a}_\mu^\dagger | \Phi_m^{N-1,RCS} \rangle$$
$$+ \sum_{I_{RCS}} c_{I_{RCS}}(t) | \tilde{\Psi}_{I_{RCS}}^N \rangle^{[n]} + c_0(t) | \Psi_0^{RCS} \rangle^{[n]}. \quad (2)$$

Here $| \Psi_0^{RCS} \rangle^{[n]}$ is the n-th order RCS correlated ground state, and $| \tilde{\Psi}_{I_{RCS}}^N \rangle^{[n]}$ indicates the excited intermediate states of the nth-order ADC(n) scheme, built using the single-particle RCS basis. The ansatz of Eq. (2) includes a full description of electron correlation effects, such as shakeup processes, breakdown of the molecular orbital (MO) picture and inter-channel couplings in the continuum, which can play an essential role both during the ionization event and the post-ionization charge dynamics.

Within the extended second-order ADC(2)x scheme for many-electron excitation used in this work, the excited (N)-electron intermediate states span the configuration space consisting of one-hole – one-particle (1h1p) $\left( | \tilde{\Psi}_{\alpha i}^N \rangle^{[n]} \right)$ and two-hole – two-particle (2h2p) $\left( | \tilde{\Psi}_{\alpha\beta ij}^N \rangle^{[n]} \right)$ excitation classes built on top of the RCS correlated ground state (24). The first term on the right-hand side of Eq. (2) describes the IS configuration states of B-spline RCS-ADC, which take the form of ionization-channel-specific product-states and reads $| \Psi_{\mu,m}^N \rangle = \hat{a}_\mu^\dagger | \Phi_m^{N-1,RCS} \rangle$, where $\hat{a}_\mu^\dagger$ is the creation operator of an electron in the IS molecular orbital $\psi_\mu(r)$, and $| \Phi_m^{N-1,RCS} \rangle$ denotes the RCS-ADC ionic eigenstates. In this work, the latter have been calculated using the ADC(2,2) (27) method for (N-1)-electron systems.

Thus, the $| \Phi_m^{N-1,RCS} \rangle$ are obtained by diagonalizing the ionic Hamiltonian calculated at the ADC(2,2) level of theory and using the single-particle RCS basis set

$$\hat{H}_{ADC(2,2)}^{N-1,RCS} | \Phi_m^{N-1,RCS} \rangle = I_m^p | \Phi_m^{N-1,RCS} \rangle, \quad (3)$$

where the ionization energy is given by $I_m^p = E_m^{N-1} - E_0^{RCS}$ and $E_0^{RCS}$ is the energy of the RCS ground state. Within the RCS-ADC(2,2) scheme, the contributions to electron correlation of one-hole (1h) and two-hole – one-particle (2h1p) configurations, is taken into account in a better-balanced way compared to both ADC(2)x and ADC(3) methods (27), making the approach more appropriate for spectral region with strong MO picture breakdown. The ionic states are expanded into 1h, 2h1p and 3h2p configurations derived from the correlated RCS ground state of the neutral molecule

$$| \Phi_m^{N-1,RCS} \rangle = \sum_i V_{i,m}^+ | \tilde{\Phi}_i^{N-1} \rangle + \sum_{\alpha ij} V_{\alpha ij,m}^+ | \tilde{\Phi}_{\alpha ij}^{N-1} \rangle + \sum_{\alpha\beta ijk} V_{\alpha ij,m}^+ | \tilde{\Phi}_{\alpha\beta ijk}^{N-1} \rangle. \quad (4)$$

The total time-dependent Hamiltonian of Eq. (1) reads

$$\hat{H}_{Pump}^N(t) = \hat{H}_{RCS-ADC}^N + \hat{D}_{RCS-ADC}^N E_{Pump}(t) - i\hat{W}, \quad (5)$$

where a complex absorbing potential (CAP) of the form $\hat{W} = \eta(r - r_{CAP})^2$ $(r \geq r_{CAP})$ is used to eliminate wave packet reflections from the boundaries of the B-spline radial grid. The laser-molecule interaction driven by the x-ray pump electric field $\left(E_{Pump}(t)\right)$ is described within the dipole approximation in the length form, and $\hat{H}_{RCS-ADC}^N$ and $\hat{D}_{RCS-ADC}^N$ are the representation of the shifted field-free Hamiltonian $\hat{\tilde{H}} = \hat{H} - E_0^{RCS}$ and the dipole operator $\hat{D}$, respectively, in the space of RCS-ADC intermediate states (see Eq. (2)). These matrices contain as sub-blocks the conventional ADC matrices computed within the RCS-based excitation space, and are further augmented with the blocks corresponding to the product-states, which describe ionization and interaction in the continuum (24-26). All the blocks of the $\hat{H}_{RCS-ADC}^N$ matrix are evaluated at the ADC(2)x level of theory, with the only exception of the Hamiltonian terms describing the energy of the (N-1)-electron ionic states, evaluated at the ADC(2,2) level of theory.

The time propagation of the unknown coefficients of the neutral many-electron wavefunction Eq. (2) is performed by means of the Arnoldi/Lanczos algorithm (25, 26). During the simulation of the pump step, we have included into the expansion of the many-electron wavefunction only the open ionization channels with energy up to the double ionization threshold (DIP) and characterized by an (ionization) spectral intensity value (see ref. (24, 25)) greater than 1%. These states will be denoted in the following as $| \Phi_{n,\text{Pump-ionized}}^{N-1,RCS} \rangle$. Doing so guarantees that all the ionic states that can be effectively populated in the x-ray pump ionization, and consequently play the main role in the ensuing many-electron dynamics taking place in the molecular cation, are accounted for (25, 26). The contribution of ionic states with a smaller spectral-intensity value, as well as of deeper core-ionization channels, was indeed found to be negligible.

The pump-triggered coherent electron dynamics is obtained by calculating, in the basis of ionic eigenstates $| \Phi_{m,\ Pump-ionised}^{N-1,RCS} \rangle$, the time-dependent reduced ionic density matrix (R-IDM) $\hat{\rho}^{R-IDM}(t)$ of the molecular ion emerging from the femtosecond pump-ionization step. This is achieved by tracing out the unobserved photoelectron degree of freedom from the total time-dependent density matrix of the N-electron system

$$\hat{\rho}^{R-IDM}(t) = \text{Tr}_\mu \left[ | \Psi^N(t) \rangle \langle \Psi^N(t) | \right]. \tag{6}$$

Doing so, and using the many-electron states of Eq. (2) within the TD RCS-ADC(n) framework, yields in the basis of ADC(2,2) ionic eigenstates

$$\hat{\rho}_{m,n}^{R-IDM}(t) = \sum_\mu c_{m\mu}(t)\left[c_{n\mu}(t)\right]^* \\ + 2e^{i(I_n^p - I_m^p)t} \int_{-\infty}^t \sum_{\mu,\nu} w_{\nu,\mu} c_{m\mu}(t')\left[c_{n\nu}(t')\right]^* e^{-i(I_n^p - I_m^p)t'} dt', \tag{7}$$

where the latter term corrects for the loss of norm introduced by the CAP (24), $I_m^p$ is the ionization potential of the ionic state m and $w_{\nu,\mu}$ is the CAP matrix element between photoelectron IS orbitals μ and ν. From now on we shall omit the R-IDM superscript and denote the reduced ionic density matrix as $\rho_{m,n}$.

The R-IDM fully describes the mixed ionic state prepared by attosecond molecular ionization: the TD population of the different ionic eigenstates is given by the diagonal elements

$$p_n(t) \,=\, |\rho_{n,n}(t)|. \tag{8}$$

while the off-diagonal ones $\rho_{m,n}$ are related to the degrees of quantum electronic coherence, $0 \leq G_{m,n} \leq 1$, between any pair of populated ionic channels m and n

$$G_{m,n}(t) \,=\, \frac{|\rho_{m,n}(t)|}{\sqrt{p_m(t) * p_n(t)}}. \tag{9}$$

In addition, the relative phases $\varphi_{m,n}$ between the partially-coherently populated eigenstates of energies $I_m^p$ and $I_n^p$ are extracted from the phases of the R-IDM off-diagonal matrix elements and read

$$\varphi_{m,n} \,=\, \arg\left[\rho_{m,n}(t)\, e^{+i(I_m^p - I_n^p)t}\right]. \tag{10}$$

The relative phase matrix is antisymmetric, i.e. $\varphi_{m,n} = -\varphi_{n,m}$, as follows from the hermiticity of the density matrix ($\rho_{n,m} = \rho_{m,n}^*$).

Finally, diagonalization of the ionic density matrix yields the so-called Schmidt decomposition (26). The latter represents a powerful tool to analyze the produced mixed state of the cationic system, allowing one to express it as a fully-incoherent sum of several ($N_{ionic}$) fully quantum-coherent pure states, each of them populated with weight $r_m$ and represented by the projection operator $\hat{P}_m$:

$$\hat{\rho}(t) \,=\, \sum_{m\,=\,1,\ N_{ionic}} r_m(t)\, \hat{P}_m(t) \,=\, \sum_{m\,=\,1,\ N_{ionic}} r_m(t)\, |\,\Phi_m^\rho(t)\,\rangle\langle\,\Phi_m^\rho(t)\,| \tag{11}$$

This operation is called purification of the density matrix.

**S1.2 Second ionization of the cationic species by the x-ray probe pulse.**

In this work, we explicitly simulate the interaction of the pump-ionized system with the probe pulse. Doing so allows us to calculate the time-delay dependence of probe-induced electron emission from the pump-ionized glycine cation, thus obtaining a realistic description of the ultrafast electronic observables measured in the experiment.

The formal validity of the presented theoretical framework relies on non-overlapping pump and probe pulses. The reason for this restriction is that, while the current B-spline RCS-ADC implementation can accurately treat many-electron dynamics with one photoelectron in the continuum, it cannot yet afford the modelling of two photoelectrons in the continuum at the same time. The latter, which would in fact be more appropriate in the case of overlapping pump and probe pulses, would also require one to extend the expansion of the many-electron RCS-ADC wavefunction (Eq. (2)) by states of the type $\hat{a}_\mu^\dagger \hat{a}_\nu^\dagger \mid \Omega_m^{N-2,RCS}\rangle$, thus posing an extra demand on the computation by means of the TD RCS-ADC machinery. However, in practise, this only limits the range pump-probe scenarios that can be numerically tackled to the ones corresponding to time-delays greater than the duration of the x-ray pulses involved $\tau_d > T_{Pump}/2$, i.e. in this case corresponding to delay times of 1.3 fs or longer.

The time-dependent description of the cation-probe interaction is performed by solving the time-dependent von Neumann equations (26) for the characterized reduced ionic density matrix (Eq. (7)) interacting with the probe laser field

$$\frac{d}{dt}\hat{\rho}(t) = -\frac{i}{\hbar}\left[\hat{H}_{Probe}^{N-1}(t),\hat{\rho}(t)\right].  \tag{12}$$

Eq. (12) is solved by using the B-spline RCS-ADC method, which we extended here to describe the photoionization dynamics of (N-1)-electron systems. The solution provides an explicit description of the ionization continua of the dication as well as of the Auger decay process, and both photoemission channels are included in the calculation. In particular, this equation can be solved by taking advantage of the Schmidt decomposition Eq. (11) of the ionic density matrix upon the pump-ionization step. It allows us to tackle the solution of Eq. (12) by propagating $N_{ionic}^{t=0}$ independent TDSE,

$$i\,\hbar\,\frac{\partial \mid \Phi_m^\rho(t)\rangle}{\partial t}\;=\;\widehat{H}_{Probe}^{N-1}(t)\;\mid \Phi_m^\rho\,(t)\rangle \quad m\;=\;1,\ldots\ldots,N_{ionic}^{t=0} \qquad (13)$$

corresponding to the individual pure ionic states obtained by the purification of the initial (before the interaction with the probe pulse) ionic density matrix prepared by the pump pulse. The time-dependent $\mid \Phi_m^\rho(t)\rangle$ states can be expanded in the basis of (N-1)-electron RCS-ADC states as

$$\mid \Phi_m^{\rho,N-1}\,(t)\rangle\;=\;\sum_{m,\mu}c_{m,\mu}(t)\;\hat{a}_\mu^\dagger \mid \Omega_m^{N-2,RCS}\rangle + \sum_{I_{RCS}}c_{I_{RCS}}(t)\mid \widetilde{\Phi}_{I_{RCS}}^{N-1}\rangle^{[n]}. \qquad (14)$$

This expression can be further simplified, and renormalized in terms of cationic eigenstates, by expressing the $\mid \widetilde{\Phi}_{I_{RCS}}^{N-1}\rangle$ RCS configurations in terms of the RCS ionic eigenstates $\mid \Phi_{n',Pump-ionized}^{N-1,RCS}\rangle$ and $\mid \Phi_{n'',Core}^{N-1,RCS}\rangle$. One derives

$$\mid \Phi_m^{\rho,N-1}\,(t)\rangle\;=\;\sum_{n,\mu}c_{m;n\mu}(t)\,\hat{a}_\mu^\dagger \mid \Omega_n^{N-2,RCS}\rangle +$$

$$+\sum_{n'}c_{m;n'}(t)\mid \Phi_{n',Pump-ionized}^{N-1,RCS}\rangle + \sum_{n''}c_{m;n''}(t)\mid \Phi_{n'',Core}^{N-1,RCS}\rangle. \qquad (15)$$

In Eq. (15) we have thus expressed the wavefunction of the time-dependent pure-state channel $\mid \Phi_m^{\rho,N-1}\,(t)\rangle$ as an expansion over the full electronic spectrum of the ionic subsystem, including both its low-energy bound excitations, i.e. the valence-ionized states populated by the pump pulse, the C(1s) core-ionized states, which were not populated in the pump-step simulation, and the electronic continua of the valence-doubly-ionized states.

The time-dependent ionic Hamiltonian $\widehat{H}_{Probe}^{N-1}(t)$ in the dipole approximation is given by

$$\widehat{H}_{Probe}^{N-1}(t) = \widehat{H}_{RCS-ADC}^{N-1} + \widehat{D}_{RCS-ADC}^{N-1}E_{Probe}(t) - \mathrm{i}\widehat{W}, \qquad (16)$$

where $\hat{H}_{RCS-ADC}^{N-1}$ and $\hat{D}_{RCS-ADC}^{N-1}$ are again the representations of the shifted field-free ionic Hamiltonian and the dipole operator, respectively, in the (N-1)-electron configuration space spanned by all the ionic states of Eq. (15). In this work, the RCS core-ionized states are calculated at the ADC(2)x level of theory employing the core-valence approximation, i.e. $\langle \Phi_{n',\text{Pump-ionized}}^{N-1,RCS} \mid \hat{H}_{RCS-ADC}^{N-1} \mid \Phi_{n'',Core}^{N-1,RCS} \rangle = 0$, and read

$$\hat{H}_{ADC(2)x}^{N-1,RCS} \mid \Phi_{n'',Core}^{N-1,RCS} \rangle = I_{n''}^{p} \mid \Phi_{n'',Core}^{N-1,RCS} \rangle \; ; \; I_{n''}^{p} = E_{n}^{N-1} - E_{0}^{RCS} \qquad (17)$$

$$\mid \Phi_{n'',Core}^{N-1,RCS} \rangle = \sum_{i} V_{i,n''}^{+} \mid \tilde{\Phi}_{i}^{N-1} \rangle + \sum_{\alpha ij} V_{\alpha ij,n''}^{+} \mid \tilde{\Phi}_{\alpha ij}^{N-1} \rangle. \qquad (18)$$

In Eq. (18) each of the ADC configurations in the expansion features at least one hole-index (i, j) correspond to either the 4a' or 5a' C(1s) occupied molecular orbital; Eq. (18) thus describes the core-ionized states of the glycine cation in the energy range of the carbon K-edge.

The first term on the right-hand side of Eq. (15) describes the IS configuration states of B-spline RCS-ADC for (N-1)-electron systems. These states take the form of ionization-channel-specific product-states

$$\mid \Phi_{\mu,n}^{N-1} \rangle = \hat{a}_{\mu}^{\dagger} \mid \Omega_{n}^{N-2,RCS} \rangle \qquad (19)$$

built upon the RCS-ADC eigenstates of the dication $\mid \Omega_{n}^{N-2,RCS} \rangle$. The latter are obtained by diagonalizing the dicationic ADC Hamiltonian $\hat{H}_{ADC(1)}^{N-2,RCS}$ computed using the single-particle RCS basis set

$$\hat{H}_{ADC(1)}^{N-2,RCS} \mid \Omega_{n}^{N-2,RCS} \rangle = DIP_{n} \mid \Omega_{n}^{N-2,RCS} \rangle, \qquad (20)$$

where the double-ionization energy is given by $DIP_{n} = E_{n}^{N-2} - E_{0}^{RCS}$. In this work, the eigenstates of the dication (produced by the probe pulse) $| \, \Omega_{n,\text{Probe-ionized}}^{N-2,RCS} \rangle = \sum_{ij} V_{ij,n}^{2+} \, | \, \widetilde{\Omega}_{ij}^{N-2} \rangle$ are described within the ADC(1) scheme for (N-2)-electron systems, and are expanded into 2h $| \, \widetilde{\Omega}_{ij}^{N-2} \rangle$ configurations derived from the correlated ground state of the neutral molecule. They are thus described up to first order in the many-body perturbation theory.

The remaining blocks $\langle \, \Phi_{n',\text{Pump-ionized}}^{N-1,RCS} \, | \, \widehat{\widetilde{H}}_{RCS-ADC}^{N-1} \, | \, \Phi_{\mu,n}^{N-1} \rangle$,

$\langle \, \Phi_{n',Core}^{N-1,RCS} \, | \, \widehat{\widetilde{H}}_{RCS-ADC}^{N-1} \, | \, \Phi_{\mu,n}^{N-1} \rangle$ and $\langle \, \Phi_{\nu,n'}^{N-1} \, | \, \widehat{\widetilde{H}}_{RCS-ADC}^{N-1} \, | \, \Phi_{\mu,n}^{N-1} \rangle$ are calculated at the ADC(2)x level of theory.

The initial condition for the solution of Eq. (13) reads

$$| \, \Phi_m^{\rho,N-1} \, (t = \tau_d) \, \rangle = \sum_{n'} c_{m;n'}(t = 0) \, e^{-\frac{i}{\hbar} I_{n'}^p \tau_d} \, | \, \Phi_{n',\text{Pump-ionized}}^{N-1,RCS} \rangle. \qquad (21)$$

It is important to note that the validity of the presented modelling of the cation-probe interaction requires the effect of the interaction with the x-ray probe pulse, on the ensuing many-electron dynamics of interest, to dominate over the residual interaction of the produced cationic system with the emitted photoelectron. For the aforementioned values of the pump-probe time-delay used here, this is indeed the case, as confirmed by the fact that the ionic density matrix $\hat{\rho}(t)$ stabilizes within the first 300 as after the pump ionization event. This is a result of the high kinetic energy of the photoelectron leaving the molecular region, and justifies the neglect of the pump-emitted photoelectron during the description of the probe step, i.e. at times greater than 1.3 fs.

Analogously to the pump-step procedure, we calculate here the time-dependent reduced dicationic density matrix (R-DIDM) $\hat{\rho}^{R-DIDM}(t)$ emerging from the probe-ionization step, by tracing out the unobserved photoelectron degree of freedom from the total time-dependent density matrix of the (N-1)-electron system

$$\hat{\rho}^{R-DIDM}(t) = \hat{\rho}^{2+} = Tr_\mu \left[| \, \Phi_m^{\rho,N-1} \, (t) \, \rangle \, \langle \, \Phi_m^{\rho,N-1} \, (t) \, | \right]. \qquad (22)$$

The density matrix of the molecular dication, in the basis of ADC(2) dicationic eigenstates $|\Omega_{m,\ Probe-ionised}^{N-2,RCS}\rangle$, reads

$$\hat{\rho}_{m,n}^{2+}(t) = \sum_{\mu} c_{m\mu}(t)[c_{n\mu}(t)]^*$$
$$+ 2e^{i(DIP_n-DIP_m)t} \int_{-\infty}^{t} \sum_{\mu,\nu} w_{\nu,\mu} c_{m\mu}(t')[c_{n\nu}(t')]^* e^{-i(DIP_n-DIP_m)t'} dt'. \quad (23)$$

Finally, the measurable photoemission spectrum resulting from the interaction of the x-ray probe pulse with the pump-prepared cationic system is recovered by convoluting the discrete, final populations of dicationic states $p_n^{2+}(t) = |\rho_{n,n}^{2+}(t)|$ with a Gaussian function of FWHM width $\delta \approx 2.5$ eV

$$p_{e^-}^{Probe-pulse}(E) = \sum_n p_n^{2+}(\infty) e^{-ln(2)\left[\left(E-I_m^p-\hbar\omega_{Probe}+DIP_n\right)/2\delta\right]^2} \quad (24)$$

The latter takes into account both the broadenings due to nuclear vibrations, probe pulse bandwidth and the instrument resolution. We also found that the position of the phase jump observed in the calculated photoemission spectra (see Section S2.2) is rather insensitive to changes in the parameter $\delta$ in the $1.5 - 3.5$ eV range.

## S2. Results.

### S2.1 First ionization of the neutral species by the x-ray pump pulse: characterization of the mixed state of the cationic system prepared by pump ionization.

Pump ionization is described by directly propagating the initial ground state of the neutral molecule with the full many-body B-spline RCS-ADC(2)x/ADC(2,2) Hamiltonian. Results are shown for the equilibrium nuclear geometry of the Gly I conformer.

In the simulation we used a multicenter B-spline basis (24, 28) characterized by a radial box radius $R_{max} = 70$, a linear grid in the centre-of-mass expansion (24) with step size h = 0.3 a.u. and maximum value of the orbital angular momentum $L_{max} = 12$. The RCS single-particle basis set we used consists of the virtual orbitals of an HF calculation performed in the cc-pVDZ GTO basis set, further truncated at the threshold energy of 2 a.u.. The number of open ionic channels included in the calculation (see Eq. (2)) and the dimension of the Hamiltonian matrix in A' symmetry space are $N_{Ionic} = 38$ and $H_{RCS-ADC}^{Maxdim} = 220500$, respectively. The values of the CAP parameters used in this calculation are $\eta = 5 \times 10^{-3}$ and $r_{CAP} = 25$ a. u. The *ab initio* simulation of the pump-step has been performed using a linearly polarized pulse, characterized by a Gaussian temporal envelope, a central frequency $\hbar\omega_{Pump} = 274$ eV, peak intensity $I_{Pump} = 6 \times 10^{15}$ W/cm$^2$ and time duration $T_{Pump} \sim 1$ fs (FWHM). Convergence of the results has been obtained by using a time-step $\Delta t = 0.5$ as and a Krylov-space dimension $K = 40$ in the Arnoldi/Lanczos time propagation.

The electronic state of the emerging ionic system has been fully characterized by computing the R-IDM. Fig. S1 shows the calculated degrees of quantum electronic coherence $G_{m,n}$ and the relative phases $\varphi_{m,n}$ between each pair of ionic eigenstates populated by the x-ray pump pulse. The final stationary populations $p_n(\infty)$ of the bound ADC(2,2)-calculated eigenstates of the Gly I cation are also shown in the vertical and horizontal side panels of Fig. S1. In the calculations presented here, all the ionic time-dependent populations $p_n(t)$ have converged to their final stationary value at $t \sim 200$ as after pump ionization. The final photoelectron spectrum resulting from the ionization of neutral glycine by the x-ray pump pulse is also plotted in Fig. S1. It is recovered by convoluting the calculated discrete, final populations of cationic states Gaussian function of FWHM width $\delta \approx 2.5$ eV, which takes into account both the broadenings due to nuclear vibrations, probe pulse bandwidth and the instrument resolution:

$$p_{e^-}^{Pump}(E) = \sum_n p_n(\infty) e^{-ln(2)\left[\left(E - \hbar\omega_{Pump} + I_n^p\right)/2\delta\right]^2} \qquad (25)$$

In Fig. S1, it is possible to identify islands of strong quantum electronic coherence close to the main diagonal (the highest degrees of generated quantum electronic coherence being $\sim 0.95$), for ionic states belonging to the 10a' (highlighted in Fig. S1) and 9a' bands. In general, a robust degree of coherence is produced between pair of states with an energy gap up to the value of pump-pulse bandwidth, while the coherence produced between different ionic eigenstates with larger energy gaps rapidly decays to very low ($< 0.2$) values. The results of our *ab initio* simulation of x-ray ionization are in very good agreement with the predictions of the sudden approximation, especially in terms of the relative phases ($\varphi_{m,n} = 0$) and relative populations of the ionic states belonging to the bands which show a breakdown of the MO picture, such as the 10a' (and 9a') inner-valence ionized band for which the sudden-approximation ansatz can be written as (using Eq. (4)) $| \Phi_{SUDDEN}^{10a',N-1} \rangle = | (10a')^{-1} \rangle = \sum_m V_{10a',m}^+ | \Phi_{m,\text{Pump-ionized}}^{N-1,RCS} \rangle$. The good accuracy of the sudden approximation for molecular ionization at x-ray photon energies is a result of the high (relative to the energy-scale of the bound electrons dynamics) kinetic energy of the photoelectron leaving the molecular region.

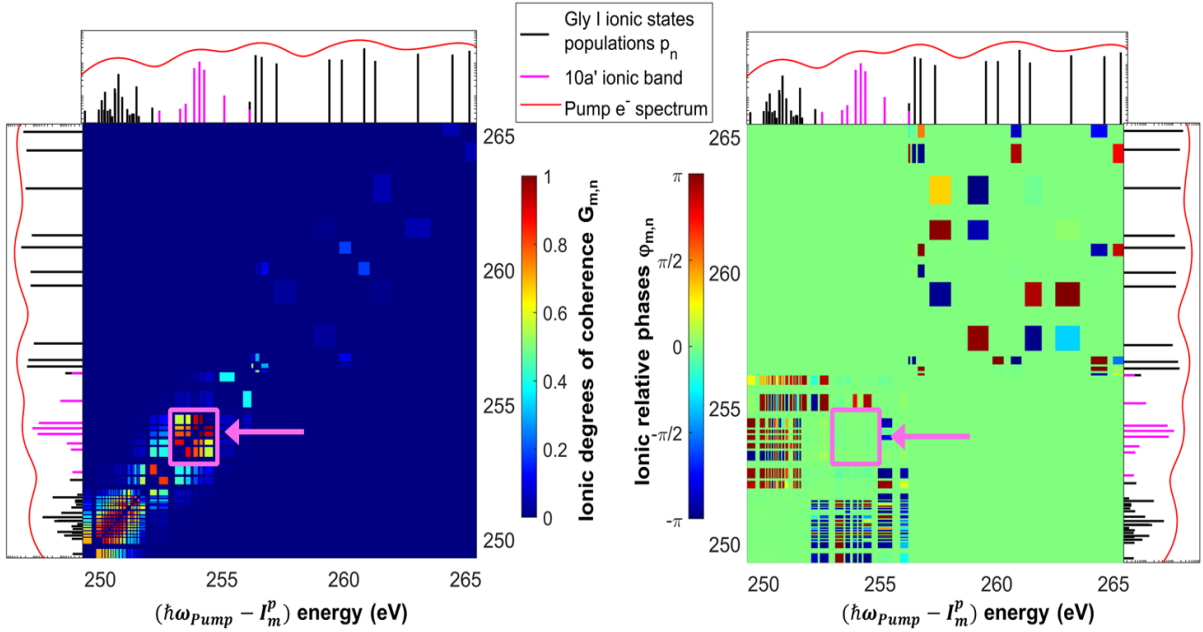

**Fig. S1.** *Ab initio* TD B-spline RCS-ADC simulation of the pump-induced first ionization of molecular glycine (Gly I conformer). The simulation is performed using the measured FLASH pulse parameters. The interaction between the pump X-ray pulse and the neutral Gly I molecule prepares a cationic system (Gly I)$^+$ in a mixed state characterized by a density matrix. The left panel shows the degrees of quantum electronic coherence $G_{m,n}$ (Eq. (9)) produced between each pair of ADC(2,2)-calculated cationic eigenstates populated by the pump pulse. The relative phases $\varphi_{m,n}$ (Eq. (10)) between each pair of ADC(2,2)-calculated ionic eigenstates are shown in the right panel. The areas corresponding to states of the 10a' band are highlighted. The populations of the ionic eigenstates, plotted against the corresponding photoelectron energy ($\hbar\omega - I_m^p$), are also shown in the vertical and horizontal side panels; the ionic states of the 10a' band are highlighted by a different stick colour. The kinetic energy distribution of pump-ionized electrons (Eq. (25)) is also shown in the vertical and horizontal side panels (red curve).

**S2.2 Second ionization of the cationic species by the x-ray probe pulse: quantum beatings in the time-resolved photoelectron kinetic-energy distribution observable.**

Results are calculated for the nuclear equilibrium geometry of the Gly I conformer. X-ray photoionization of the cationic system by the probe pulse is described here by solving Eqs. (13) and (15) for the cases of the three principal quantum-coherent pure-state channels, denoted here as $\mid \Phi_1^\rho(t)\rangle$, $\mid \Phi_2^\rho(t)\rangle$, $\mid \Phi_3^\rho(t)\rangle$, as obtained in Eq. (11) by the purification of the ionic density matrix prepared by the pump pulse. These principal, coherent ionic channels correspond to the coherent superpositions of ionic states in the 10a' ($I_{10a'}^p \sim 20$ eV, $r_1 = 0.15$) and 9a' ($I_{9a'}^p \sim 23$ eV, $r_3 = 0.09$) bands, respectively, as well as the coherent superposition of the two ionic states consisting of 11a' and 12a' hole-mixing ($r_2 = 0.13$) at an energy $I_{11a'/12a'}^p \sim 17.5$ eV.

The absolute value of the expansion coefficients of these pure-state channels (in the basis of the 38 ionic eigenstates $\left\{\mid \Phi_{m,\text{Pump}-\text{ionized}}^{N-1,RCS}\rangle\right\}$ included in the simulation of the x-ray pump ionization step) as calculated by Eq. (11), are shown in the insets of Fig. S3. The other pure-state channels contributing to the expansion of Eq. (11) are characterized by a negligible quantum coherence in the basis of the cationic eigenstates and therefore do not contribute to the observed oscillations in the photoelectron observables.

In order to illustrate the charge dynamics triggered by the x-ray pump ionization, in Fig. S2 we show the hole densities (defined as the difference between the electron density of the neutral ground state and the one of the pump-prepared correlated ionic state), corresponding to the 10a' band coherent ionic channel $\mid \Phi_1^\rho(t)\rangle$, at two different times, namely immediately after the pump ionization event and 11 fs after it.

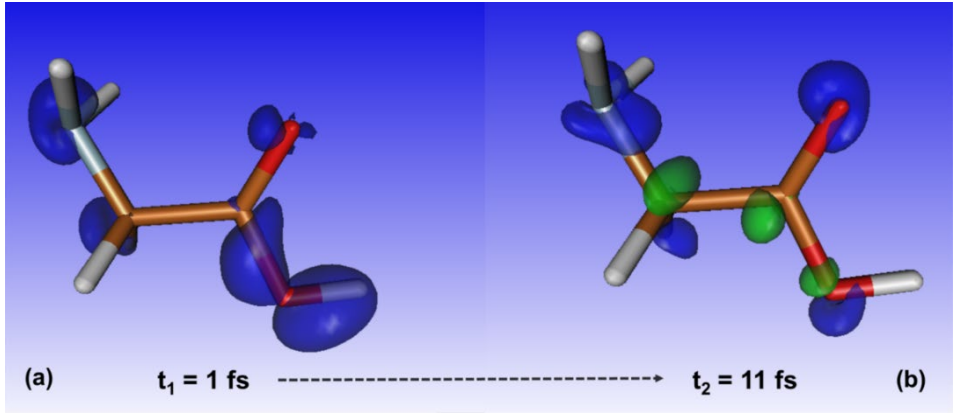

**Fig. S2. Hole densities corresponding to the correlated 10a' pure state channel at $t_1$ = 1 fs (a) and $t_2$ = 11 fs (b) after the pump ionization event.** The density iso-surfaces displayed are the ones with value 0.015, blue and green colors indicate positive and negative values of the hole density, respectively.

The time propagation is performed by means of the Arnoldi/Lanczos algorithm. During the simulation of the probe step, we have included in the expansion of the many-electron wavefunction Eq. (15) all the open doubly-ionized channels with energy up to 3.5 a.u. The resulting number of open dicationic channels included in the simulation is $N_{Dication} = 225$. The highest value (among the different molecular symmetry spaces) of the Hamiltonian matrix dimension is $H_{RCS-ADC}^{Maxdim} = 1335318$. The other numerical parameters are the same as for the simulation of the pump step (see Section S2.1).

To reduce the demanding computational effort resulting from the size of the Hamiltonian matrix, we employed the following approximation and neglected the interchannel couplings in the continuum between different dicationic channels, i.e. we set

$$\langle \Phi_{\nu,n'}^{N-1} \mid \hat{H}_{RCS-ADC}^{N-1} \mid \Phi_{\mu,n}^{N-1} \rangle = 0, \ n' \neq n.$$

Fig. S3 shows the relative variation, as a function of the x-ray pump-probe time-delay $\tau_d$, of the separate contributions to the probe-induced electron yield $p_{e^-}^{Probe-pulse}$ (Eq. (24)) from the three pure-state ionic coherent superpositions associated with the 10a', 11a'/12a' and 9a' bands, respectively. The change of electron yield is plotted relative to the time-delay averaged value. Since the time-delay oscillation of the relative change in $p_{e^-}^{Probe-pulse}(E, \tau_d)$ shows, for each coherent ionic channel, a change of phase by $\pi$ in different regions of the kinetic energy spectrum, we plot the integrated $p_{e^-}^{Probe-pulse}(\tau_d)$ yield in both the kinetic energy ranges above and below the phase jump position resulting from the *ab initio* numerical calculations.

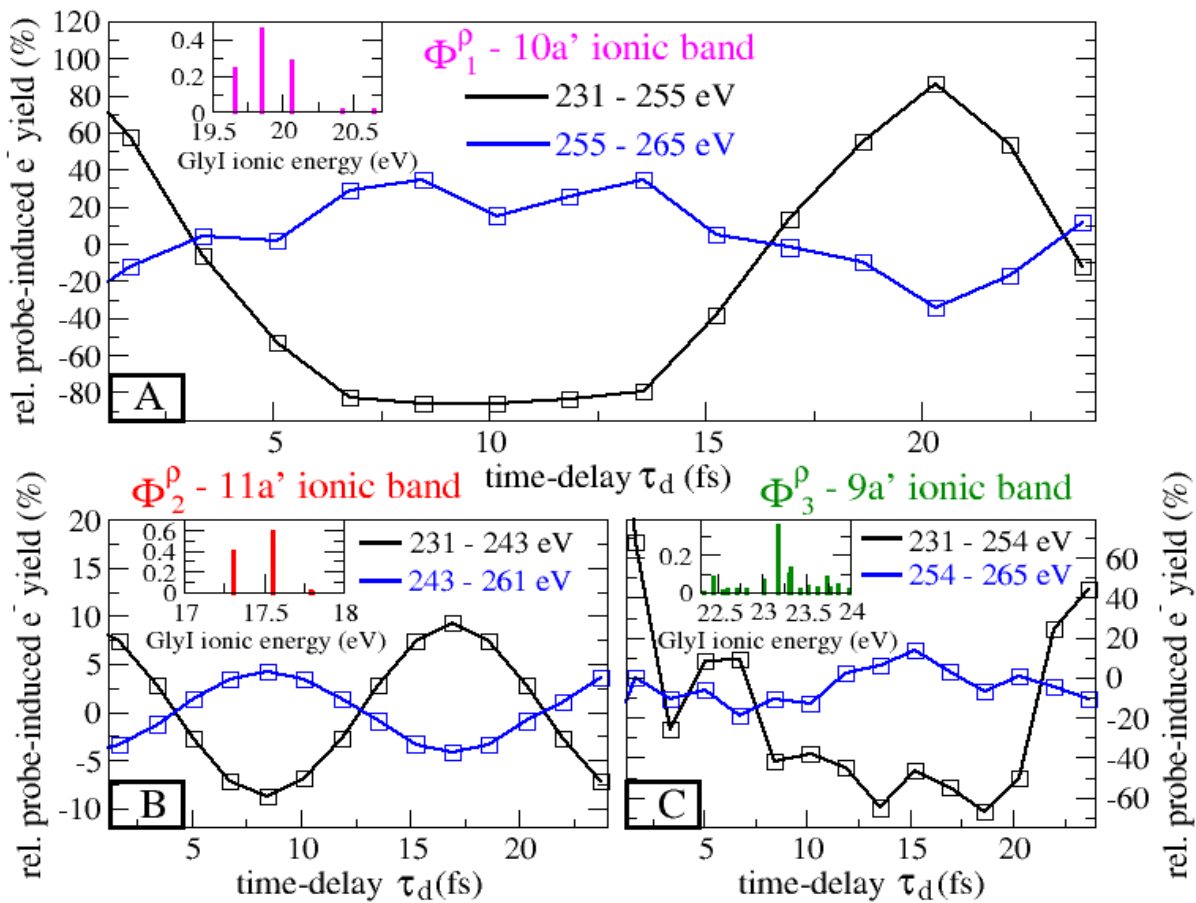

**Fig. S3.** *Ab initio* **TD B-spline RCS-ADC simulation of the interaction between the probe pulse and the pump-prepared cationic system (Gly I conformer). The simulation is performed using the measured FLASH pulse parameters. (A)** Relative change of the probe-produced electron yield in the kinetic energy ranges above and below the phase jump, as a function of x-ray pump-probe delay; contribution from the pump-prepared pure-state ionic channel consisting of a coherent superposition of eigenstates in the 10a' ionic band. In both energy ranges the electron yield includes contributions from Auger emission and sequential double ionization as a function of x-ray pump-probe delay. **(B)** and **(C)** – same as **(A)** for the contribution of the pump-prepared pure-state ionic channels consisting of coherent superpositions of eigenstates in the 11a'/12a' and 9a' ionic bands, respectively.

The calculated position of the phase jump for the 10a' and 9a' coherent ionic channels (~255 eV) appears shifted to a higher value of the electron kinetic energy with respect to the one observed in the experimental data. This deviation shows that, despite the excellent agreement in capturing at a qualitative level the phase-jump mechanism, the theoretical modelling of the final doubly-ionized states $| \Omega_{n,Probe-ionized}^{N-2,RCS} \rangle$ adopted here does not accurately include the effect of electron correlation in the populated final states of the glycine dication. The part of the spectrum where the phase jump is observed in the experiment is a region of complete breakdown of the MO picture of double ionization, characterized by a high density of doubly-ionized satellite states. As a result, the contribution of 3h1p configurations to the description of the states in this energy region is very strong. Since this contribution is not included here (at the ADC(1) level, 2h configuration mixing is taken into account), the absolute position of the dicationic energy band is not well-captured at a quantitative level, leading to a shift of the predicted phase-jump position.

In both energy ranges the electron yield includes contributions from both the direct sequential double ionization (SDI) by the probe pulse of the cationic states populated coherently by the pump pulse, and the Auger emission from the probe-excited 4a' and 5a' core-ionized Auger channels.

In order to quantify the relative contribution of these two mechanisms, in Fig. S4 we show the relative change, as a function of the x-ray pump-probe time-delay, of the total contribution from the pump-prepared cationic mixed state to probe-induced electron yield in the kinetic energy ranges above and below the phase jump. For comparison, we plot also the isolated contribution due to electrons emitted through Auger-decay. The total spectrum shown in Fig. S4 consists of the incoherent sum of the individual contributions from the 3 coherent pure-state channels shown in Fig. S3. The Auger electron contribution for each coherent channel is calculated as

$$p_{e^-}{}^{Auger}(E) = \sum_n p_n^{2+,Auger}(\infty)\, e^{-ln(2)\left[\left(E-I_m^p-\hbar\omega_{Probe}+DIP_n\right)\Big/2\delta\right]^2} \qquad (26)$$

by integrating the CAP term of the diagonal dicationic density matrix (Eq. (23)) starting from $t = 0.5$ fs after the probe pulse

$$p_n^{2+,Auger}(\infty) = \left| \rho_{n,n}^{2+,Auger}(\infty) \right| = \left| \lim_{t \to \infty} 2 \int_{0.5fs}^{t} \sum_{\mu,\nu} w_{\nu,\mu} c_{m\mu}(t')[c_{n\nu}(t')]^* \, dt' \right| \quad (27)$$

The validity of this formula is based on the different time-scales corresponding to direct and Auger electron emission, and by the fact that the photoelectron wave-packet emitted as a result of direct photoionization by the probe pulse is completely absorbed by the CAP within the first 0.3 fs after the probe pulse. The total Auger electron contribution presented in Fig. S4 thus describes the total yield of electrons emitted in the Auger decay of the C(1s) 4a' and 5a' core-ionized states populated by the probe pulse.

The result of Fig. S4 show that, at the photon energy value used in the present experiment, the direct photoionization dominates over the excitation of the intermediate Auger-active core-singly-ionized C(1s) resonances. The calculated Auger signal is indeed about one order of magnitude smaller than the SDI one.

Therefore, the quantum-coherent electron dynamics prepared by the x-ray pump in the cation is imprinted in the direct probe-ionized electron signal, which dominates over the Auger-induced electron signal. Moreover, the results show that the photoelectron signal is more sensitive to the quantum coherence in the 10a' band compared to the coherences produced in the 11a'/12a' and 9a' ionic bands: this is because the pure-state ionic channel consisting of a coherent superposition of eigenstates in the 10a' band is characterized by a higher population weight $r_{10a'} = 0.15$ (see Schmidt decomposition in Eq.(11)), and, as shown in Fig. S3, also gives rise to higher-amplitudes in the calculated oscillation of the electron yield. As a consequence, the position of the phase jump in the total time-delay dependent electron yield is also approximately the same as the one of the 10a' coherent channel alone.

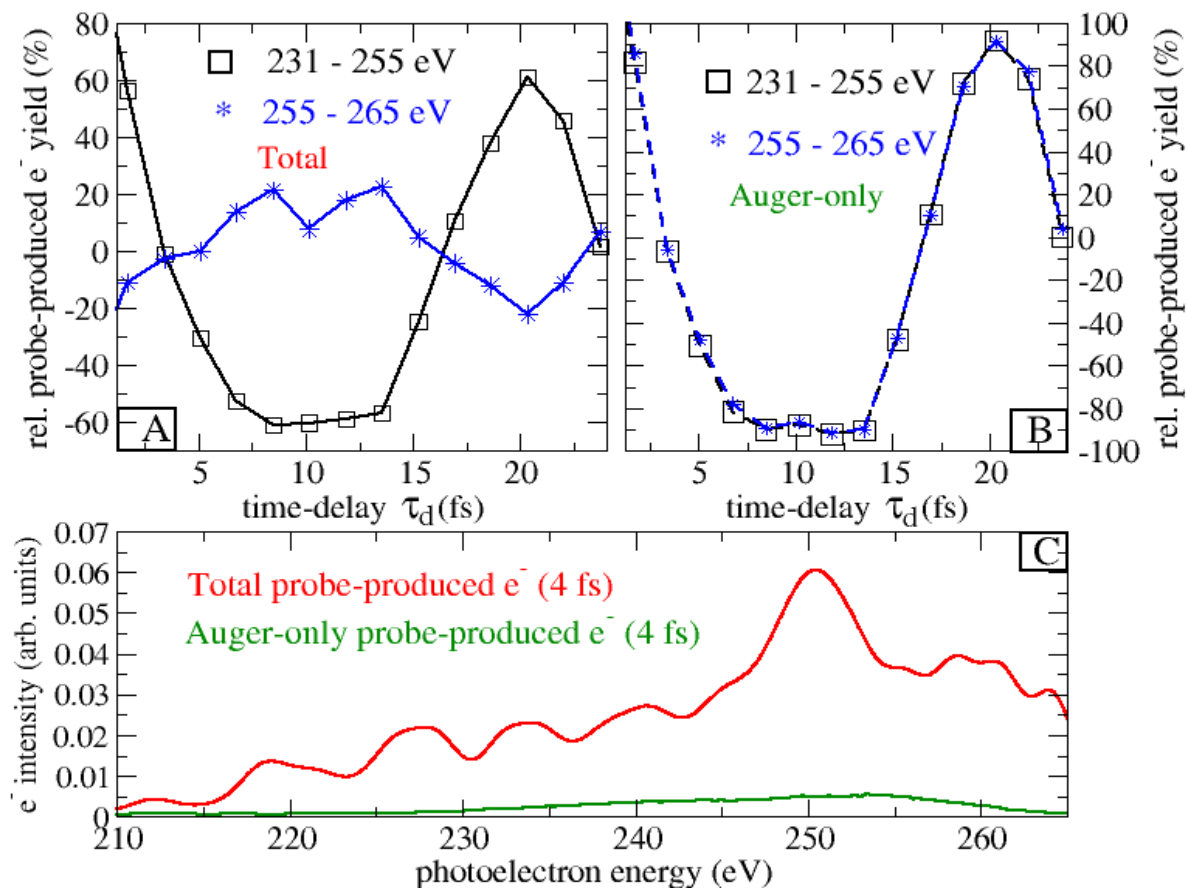

**Fig. S4.** *Ab initio* **TD B-spline RCS-ADC simulation of the interaction between the probe pulse and the pump-prepared cationic system (Gly I conformer). The simulation is performed using the measured FLASH pulse parameters. (A)** Relative change of the probe-produced electron yield in the kinetic energy ranges above and below the phase jump, as a function of x-ray pump-probe delay; total contribution from the pump-prepared mixed-state of the ionic system. In both energy ranges the electron yield includes contributions from Auger emission and sequential double ionization as a function of x-ray pump-probe delay. **(B)** same as **(A)** for the isolated contribution of Auger electrons, resulting from the decay of the 5a' and 4a' core-ionized states populated by the probe. **(C)** Kinetic energy distributions of the probe-induced Auger electrons (green curve) and the complete (Auger + SDI) probe-emitted electrons (red curve); the latter includes the dominant contribution of the photoelectron emitted in the process of (second) direct photoionization of the cationic system by the probe pulse.

## S3. Effect of nuclear ground-state geometry distribution: time-dependent survival probability of the pump-prepared mixed state.

Here we analyze the effect of the geometry-spread of the ground state nuclear wavefunction on the survival of the quantum electronic coherences in the 10a' and 9a' ionized bands. The simulations were performed for and averaged over 201 different nuclear geometries, at which convergence of the calculated ionic density matrix survival probability was found.

**S3.1 Numerical procedure.**

The entire procedure of the calculation is outlined below. For each and every nuclear geometry, the calculation we performed consists of three elements:

**1-** Calculation of the bound ionic states of glycine, in the 10a' and 9a' bands energy regions. This calculation was performed at the advanced ADC(2,2) level of theory (27) by using the cc-pVDZ basis set with the virtual orbital space truncated at the threshold energy of 2 a.u. Here, only ionic eigenstates with more than 1% 10a' or 9a' contribution in their full ADC configuration expansion are kept.

**2-** The pump-prepared initial mixed-state of the cationic system, characterized by the ionic density matrix $\hat{\rho}(t_0)$, is time-propagated by solving the (field-free) time-dependent von Neumann equation (26)

$$\frac{d}{dt}\,\hat{\rho}(t) \,=\, -\frac{i}{\hbar}\big[\hat{H}_{ADC(2,2)}^{N-1,RCS}\,,\hat{\rho}(t)\big], \tag{28}$$

$$\hat{\rho}(t) \,=\, e^{-\frac{i}{\hbar}\,\hat{H}_{ADC(2,2)}^{N-1,RCS}\,(t-t_0)}\,\hat{\rho}(t_0)\,\exp^{+\frac{i}{\hbar}\,\hat{H}_{ADC(2,2)}^{N-1,RCS}\,(t-t_0)}. \tag{29}$$

For each nuclear geometry, the initial $\hat{\rho}(t_0)$ reduced ionic density matrix is built according to the following criteria, which we name *coherence-corrected sudden-approximation ansatz*:

- The relative populations of the valence ionic eigenstates (both in the 10a' and 9a' bands), as well as the relative phases between each pair of eigenstates, are estimated using the sudden approximation ansatz to model the ionization of neutral glycine by the pump pulse. Therefore, each ionic eigenstate has a population proportional to the weight of the simple one-hole configurations in their ADC configuration expansion, and the relative phases are 0.

- We assume a degree of quantum coherence different from zero only between ionic eigenstates of the same symmetry, and consequently do not include ionic states of a'' symmetry in our description. Assuming a coherence different from zero only between ionic states of the same symmetry implies that the direction of the first emitted photoelectron is either not measured or integrated out in the analysis of the experimental data. At a theoretical level, it corresponds to calculating the ionic density matrix by tracing the full N-electron neutral wavefunction over the spatial-symmetry quantum numbers of the photoelectron.

- Finally, for each pair of coherently populated ionic eigenstates, the degree of coherence is further decreased (from the initial value of 1) in order to take into account the effect of their energy separation, the bandwidth and profile of the pump pulse. Therefore, the off-diagonal matrix elements $\rho_{m,n}(t_0)$ of the ionic density matrix are modified according to

$$\rho_{m,n}(t_0) = \rho_{m,n}^{Sudden} \, e^{-\ln(2)\left(\frac{E_m - E_n}{E_{Band}}\right)^2},\tag{30}$$

which provides a reliable estimate of the initial ionic density matrix prepared as a result of ionization by a 274 eV x-ray Gaussian pump pulse of bandwidth $E_{Band}$.

The validity of this ansatz is confirmed, at the nuclear equilibrium geometry, by the *ab initio* results of Fig. S1.

**3-** Given the density matrix $\hat{\rho}(t)$, the survival probability of the time-dependent quantum state is calculated as the fidelity between the quantum states described by the density matrix $\hat{\rho}(t)$ and the initial density matrix prepared by the pump $\hat{\rho}(t_0)$

$$\mathrm{F}(t) = \left( Tr \left[ \sqrt{\sqrt{\hat{\rho}(t)} \, \hat{\rho}(t_0) \sqrt{\hat{\rho}(t)}} \, \right] \right)^2.\tag{31}$$

## S3.2 Nuclear geometry sampling.

The optimization of the equilibrium geometry $\left\{\vec{R_n}\right\}_0$ and the calculation of the normal-mode frequencies were computed at the coupled-cluster singles and doubles (CCSD) level of theory in a cc-pVTZ basis set, using the MOLPRO quantum chemistry package (30); the different nuclear geometries were sampled in the configuration space according to a probability distribution given by the Wigner distribution, integrated over momenta, corresponding to the nuclear ground-state (GS) wavefunction, i.e. $\left|\Lambda_{GS}^{Nuclear}\left(\{\overrightarrow{R_n}\}\right)\right|^2$. The latter was calculated within the harmonic oscillator approximation (around the equilibrium geometry) of the potential energy surfaces. Therefore, the ensemble of geometries was calculated using a product of normal Gaussian distributions, each one corresponding to the respective vibrational normal mode considered. Only vibrational normal modes that conserve the molecular symmetry of the equilibrium geometry ($C_s$ point group) were considered. Monte Carlo integration of the time-dependent survival probability curves over different nuclear geometries $\{\overrightarrow{R_n}\}$ was performed. The aforementioned procedure describes the effect of the spread of the GS nuclear wavefunction on the observed coherent electron dynamics. Vibrational dynamics is not included in the simulations.

### S3.3 Results.

Fig. S5 shows the geometry-averaged, time-delay-dependent survival probability of the ionic density matrix prepared by the x-ray pump ionization. Results are presented for the two conformers Gly I and Gly III, as well as for three different statistical mixings of the populations of the two conformers in the sample.

The results for both the singly-ionized Gly I conformer and the averaged abundance-weighted contribution of the two (Gly I and Gly III) conformers, show that it is possible to discern an oscillation with a $\sim 20$ fs period in the time-dependent survival probability of the cationic mixed state prepared by the pump ionization. The overall agreement with the experimental results is excellent in terms of the oscillation period and it provides strong evidence in support of the presence of an electronic coherence-driven oscillation characterized by a period of $\sim 20$ fs and whose coherence partially survives the nuclear GS distribution averaging.

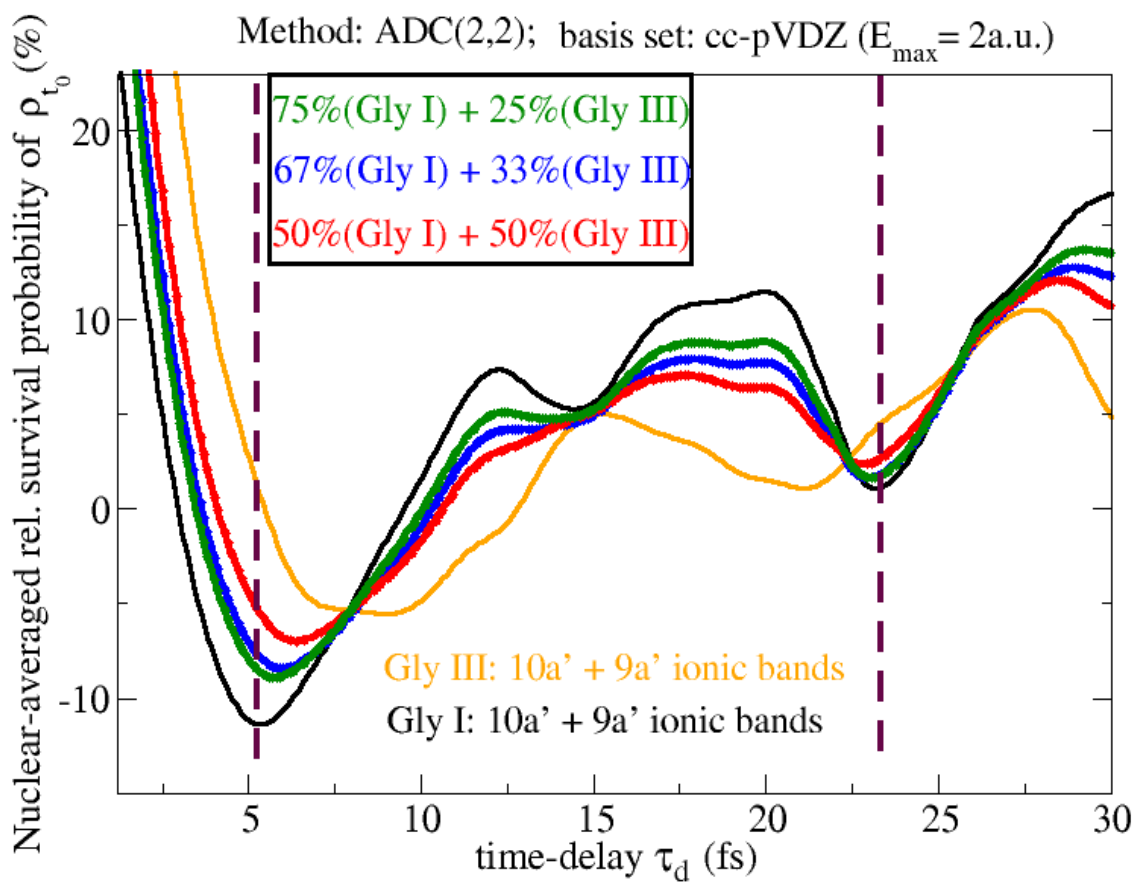

**Fig. S5. Time-dependent survival probability of the pump-prepared ionic density matrix** in the subspace of ionic eigenstates consisting of the 10a' and 9a' ionic bands. The result of Eq. (31) has been averaged over the nuclear ground-state geometry distribution for both the Gly I and Gly III conformers, as well as for three different statistical mixtures of the two.

Here it is important to note that, whereas both the time-dependence and the temporal resolution of the measured physical observable does in principle present deviations from the calculated survival probability, the robustness of the latter with respect to the nuclear geometry averaging provides a probe-free demonstration of the survival of quantum electronic coherence on the tens-of-femtosecond time-scale in this system.

Regarding the profile of the oscillation, neither the theory results nor the bare experimental data points show a simple, perfectly singled-period oscillation. Moreover, the amplitude of the calculated oscillation is around 10% (measured against the time-delay averaged value), and therefore smaller than the experimentally measured one which ranges from 15% up to 30%.

The observed difference in the relative amplitude of the oscillations has to be considered remarkably small, and thus supportive of the presented experiment's interpretation, especially considering the potential theoretical sources of discrepancy (in addition to the experimental errors), which include:

- Errors due to the use of a truncated single-particle basis set in the calculation of the 10a' and 9a' ionized states at the ADC(2,2) level of the ADC hierarchy. This can in principle have an appreciable impact on the observed discrepancy, as it can lead to relative errors in the excited ionic states expansions and consequently in the energy gaps between different ionic eigenstates as well as in the geometry of the corresponding potential energy surfaces.
- Deviations, as a function of the nuclear geometry, of the relative ionic populations and ionic coherences, estimated by means of the procedure outlined above, with respect to the pump-prepared ones. The vertical ionization probabilities by the pump could potentially present deviations from the theoretically estimated one for the states involved, depending on the vertical ionization energies at different geometries.
- The effect of phenomena not-included in the present analysis:
    i. the symmetry-breaking vibrational normal modes.
       the ionization-induced nuclear motion.
    ii. Given the relatively high temperature of the molecular sample used in the experiment, contributions from excited vibrational states could also play a non-negligible role.